\documentclass[twocolumn,trackchanges]{aastex631}

\newcommand\chandra{{\it Chandra}}

\newcommand\xmm{{\it XMM-Newton}}

\newcommand\asat{{\it AstroSat}}
\newcommand\astrosat{{\it AstroSat}}
\newcommand\nustar{{\it NuSTAR}}

\newcommand{\opt}{\texttt{OPTXAGNF}}

\newcommand{\fagn}{\texttt{FAGNSED}}
\newcommand{\dbb}{\texttt{DISKBB}}
\newcommand{\nthc}{\texttt{NTHCOMP}}

\newcommand{\zbb}{\texttt{ZBBODY}}

\newcommand{\zpl}{\texttt{ZPOWERLAW}}
\newcommand{\xab}{\texttt{XABS}}

\newcommand{\cmsq}{$\rm cm^{-2}$}
\newcommand\kms{{$\rm~km~s^{-1}$}}
\newcommand\ergA{{$\rm erg~cm^{-2}~s^{-1}$\AA$^{-1}$}}
\newcommand\ergS{{$\rm erg~cm^{-2}~s^{-1}$}}
\newcommand\Lum{{$\rm erg~s^{-1}$}}
\newcommand\ang{{$\rm \AA$}}
\newcommand\msun{{$\rm~M_{\odot}$}}
\newcolumntype{H}{>{\setbox0=\hbox\bgroup}c<{\egroup}@{}}

\newcommand\kev{{\rm~keV}}

\newcommand\ev{{\rm~eV}}
\newcommand\angs{{\rm \AA}}
\usepackage{booktabs, makecell}
\usepackage{booktabs}
\usepackage{array}
\usepackage{verbatim}
\usepackage{amsmath}
\usepackage{multirow}
\usepackage{array}
\usepackage{enumitem}
\usepackage{hyperref}
\newcolumntype{H}{>{\setbox0=\hbox\bgroup}c<{\egroup}@{}}
\usepackage{blindtext}
\usepackage{soul}
\usepackage{cancel}
\graphicspath{{./}{figures/}}
\usepackage{enumitem}
\usepackage{booktabs}
\usepackage{multirow}
\usepackage{array}
\newcolumntype{H}{>{\setbox0=\hbox\bgroup}c<{\egroup}@{}}
\usepackage{booktabs, makecell}
\usepackage{booktabs}
\usepackage{array}
\usepackage{verbatim}
\usepackage{amsmath}

\newcommand{\rev}[1]{\textbf{{\color{blue} #1}}}

\begin{document}

\title{Spectral connection between far UV and soft X-ray emission from active galactic nuclei using AstroSat}

\correspondingauthor{Shrabani Kumar}
\email{kmrshrab@gmail.com}

\author{Shrabani Kumar }
\affiliation{Inter-University Centre for Astronomy and Astrophysics, Pune, 411007, India}

\author[0000-0003-1589-2075]{G. C. Dewangan}
\affiliation{Inter-University Centre for Astronomy and Astrophysics, Pune, 411007, India}
\author[0000-0001-6952-3887]{K. P. Singh}
\affiliation{Indian Institute of Science Education and Research Mohali, Knowledge City, Sector 81, Manauli P.O., SAS Nagar, 140306, Punjab, India}
\affiliation{Department of Astronomy and Astrophysics, Tata Institute of Fundamental Research, 1 Homi Bhabha Road, Mumbai 400005, India}
\author[0000-0003-3105-2615]{P. Gandhi}
\affiliation{School of Physics \& Astronomy, University of Southampton, Highfield SO17 1BJ, UK}
%\author{I. E. Papadakis}
%\affiliation{Department of Physics and Institute of Theoretical and Computational Physics, University of Crete, 71003 Heraklion, Greece}
%\affiliation{Institute of Astrophysics—FORTH, N. Plastira 100, 70013 Vassilika Vouton, Greece}

\author[0000-0001-8624-9162]{L. Mallick}
\altaffiliation{CITA National Fellow}
\affiliation{Department of Physics \& Astronomy, University of Manitoba, Winnipeg, Manitoba R3T 2N2, Canada}
\affiliation{Canadian Institute for Theoretical Astrophysics, University of Toronto, 60 St George Street, Toronto, Ontario M5S 3H8, Canada}

\author[0000-0001-9097-6573]{G. C. Stewart}
\affiliation{Department of Physics and Astronomy, The University of Leicester, University Road, Leicester LE1 7RH, UK}

\author[0000-0002-6351-5808]{S. Bhattacharyya}
\affiliation{Department of Astronomy and Astrophysics, Tata Institute of Fundamental Research, 1 Homi Bhabha Road, Mumbai 400005, India}

\author[0000-0002-8776-1835]{S. Chandra}
\affiliation{Center for Space Research, North-West University, Potchefstroom 2520, South Africa}

\begin{abstract}

We present the UV/X-ray joint spectral analyses of four Seyfert~1 galaxies (PG~0804+761, NGC~7469, SWIFT~J1921.1-5842, and SWIFT~J1835.0+3240) using the data acquired with the Ultraviolet Imaging Telescope and Soft X-ray Telescope onboard \textit{AstroSat}. We model the intrinsic UV/X-ray continuum with the accretion disk, warm and hot Comptonization using the \opt{} and \fagn{} models, where the disk seed photons are Comptonized in the warm and hot corona. The Eddington ratio of the four Seyferts ranges from $0.01$ to $1$.  %We find that more than half ($\sim $ 60\%)  of the bolometric luminosity is contributed by the disk in the case of NGC~7469. 
In the case of SWIFT~J1835.0$+$3240, we infer a compact warm corona ($ R_{warm} - R_{hot} \lesssim 18 r_g$) while, PG~0804+761, NGC~7469, and SWIFT~J1921.1-5842 may exhibit a larger warm Comptonizing region ($\gtrsim 32r_g$). 
We could constrain the spin parameter in PG~0804+761, $a^\star = 0.76_{-0.20}^{+0.08}$ (1$\sigma$ error), with the \fagn{} model. 
%In SWIFT~J1921.1-5842, the standard disk exists beyond $\sim 95 ~r_g$  while the inner disk is converted into a warm Comptonizing plasma. In NGC~7469, the soft X-ray excess emission may favor the warm Comptonization of disk seed photons. 
In SWIFT~J1835.0+3240 and SWIFT~J1921.1-5842, the UV/X-ray spectral variability may be driven by the thermal Comptonization of the disk seed photons in the hot corona. Furthermore, the observed spectral hardening with the decrease in disk temperature and accretion rate compared to earlier observations may indicate a state transition in SWIFT~J1835.0+3240 from a high/soft to a low/hard state. 
%In PG~0804+761, the X-ray reflection from the disk could be largely responsible for the observed soft X-ray excess, as the inner disk radius extends down to $\sim 1.5~r_g$.

\end{abstract}

%\keywords{ --- Ultraviolet astronomy(1736) --- }

\section{Introduction} \label{sec:intro}

The primary emission from radio-quiet active galactic nuclei (AGN) consists of the Big Blue Bump (BBB), the soft X-ray excess component, and the broadband X-ray power-law emission. The BBB emission generally peaks in the extreme UV band and spans over near-infrared to extreme UV bands \citep{koratkar1999ultraviolet}.  This component is generally contaminated with broad/narrow emission lines, narrow absorption lines, and emission from the host galaxy and can be reddened due to the host galaxy. The BBB component is thought to be the direct consequence of the accretion flow arising from the accretion disk around central super-massive black holes in AGN \citep{1973A&A....24..337S}. However, the observed UV continua are generally found to be redder than the theoretical accretion disk spectrum. Using far UV spectra of 8 bright Seyfert 1 galaxies acquired with \textit{AstroSat} observations, we showed that the observed spectra are generally consistent with standard disk models, but the disks appear truncated (\citealt{kumar2023far}, hereafter paper~I). It is unclear if the apparent truncation of disks is real or just due to the deficit of UV emission from the innermost accretion disks. In this aspect, it is important to understand the connection of disk emission with the soft X-ray emission.

The X-ray spectra of many Seyfert 1 galaxies show the presence of soft excess components. First observed by \citet{singh1985observations} in the \textit{HEAO-1} data and \citet{10.1093/mnras/217.1.105} in the \textit{EXOSAT} data,  the soft excess is identified as an excess over the broadband X-ray power-law continuum in the soft X-ray band below $2\kev$. The temperature of this blackbody-like component is found to be remarkably similar, around $\sim 0.1$ keV across AGN with different black hole masses \citep{gierlinski2004soft,mallick2022high}. In some AGN, the short time scale variability of the soft X-ray excess emission suggests that this component arises from the innermost regions. 
%the emitting region could be closer to the accretion disk. \citet{boller1996soft} found a correlation between H$\beta$ line width and the soft X-ray excess, indicating the ionized gas in the BLR could be connected to the soft excess emission.  
The exact nature and origin of the soft excess still remain uncertain. Though several models have been proposed to explain this emission component, currently, two competing models, warm Comptonization and blurred reflection, can both explain the origin of the soft excess. X-ray reflection from a partially ionized accretion disk can give rise to many emission lines and a Thomson scattered continuum. The relativistic blurring due to special and general relativistic effects on the numerous emission lines and the scattering continuum can give rise to a smooth continuum component that mimics the soft excess component \citep{george1991x,garcia2014improved}. However, the blurred reflection spectra inferred from the broad iron line when extrapolated to the soft X-ray band below $2\kev$ appears to be insufficient for the observed strong soft excess in some AGN (e.g., Ark~120; \citealt{2017MNRAS.472..174M}). This problem can be alleviated by high-density reflection models \citep[e.g.,][]{10.1093/mnras/stw1696,2018MNRAS.479..615M}.   
Another popular model for the soft excess is the warm Comptonization model.  In this case, the soft excess component is treated as a different continuum component. This is believed to originate from a warm plasma ($kT_{w} \sim 0.1-1$ keV) with large optical depth ($\tau \sim 10-40$) in the inner region of a truncated accretion disk \citep{2018A&A...611A..59P}. The outer area of the accretion disk may still behave as a standard accretion disk. These warm layers of plasma Compton up-scatter the disk seed photons, giving rise to the apparent soft excess \citep{done2012intrinsic,kubota2018physical}. The only difficulty in this model is fine-tuning the heating and cooling of the warm corona to obtain the fixed temperature observed for a wide range of black hole masses.

These two models, blurred reflection and warm Comptonization, often produce statistically equivalent results, making it difficult to distinguish between them \citep{2007ApJ...671.1284D,pal2016x,waddell2019multi,chen2025uv}. \citet{middei2020soft} studied the narrow line Seyfert1 Mrk~359 using \textit{XMM-Newton -- NuStar} observations. They tested both the relativistic blurred reflection and warm Comptonization model and found that the latter reproduced the soft excess better. Similarly, for Zw~229.015, \citet{tripathi2019nature} observed the warm Comptonization to describe the soft excess better than other models. \citet{noda2011suzaku} found the soft X-ray variability does not follow the fast variability observed in the hard X-ray for Mrk~509. If X-ray reflection is the origin (or partial origin) for soft X-ray excess, then a correlation between the soft and hard X-ray variability is expected \citep{boissay2014multiwavelength,2018MNRAS.479..615M}. In the warm Comptonization model, since the warm corona is either the innermost part of the accretion disk or the warm layer on it, such warm coronae can modify the accretion disk substantially. Hence, it is important to study the connection between the accretion disk UV emission and soft excess emission.

In this paper, we extend our work presented in paper~I on far UV spectroscopy of AGN  and include soft X-ray data acquired simultaneously with \astrosat{}. We perform joint spectral analysis of far UV  and the soft X-ray data on four AGN and study the spectral connection between the accretion disk and the soft X-ray excess. 
%Although we lack high energy (above 7 keV) spectral data, which is the limitation in this analysis, the advantage of grating data in the far UV and near UV band will provide better constraints to describe the accretion disk emission and thereby the connection between other emission components. 
%This is an extension of the work done in \citet{kumar2023far} (hereafter paper1). 
%Due to quality issues in the X-ray data, we did not include three AGN (Mrk~841, Mrk~352, and I~Zw~1) from paper1. MR~2251-178 is also excluded due to the presence of a variable warm absorber, as the low-resolution SXT spectra may not be able to estimate the continuum accurately. 
 The broadband SED of PG0804 is presented for the first time here. 
 %Three sources are Seyfert~1 and the SWIFT1835 is a broad-line radio galaxy (BLRG) whose radio-quiet counterpart is a Seyfert 1. 
%SWIFT1835 shows FR II morphology. Typically, the optical/UV spectra of BLRG show broad emission lines, and the X-ray spectra show weak Fe~K$\alpha$ and reflection features. 
This paper is organized as follows. We describe the observations and data reduction in Section~\ref{sec:observation}, and perform joint spectral analysis in Section~\ref{sec:spec_analys}. We discuss our results in Section~\ref{sec:results} followed by a summary in Section~\ref{sec:conclusion}.

\section{Observations and Data Reduction}
\label{sec:observation}
We utilized the simultaneously acquired UV and X-ray spectral data from \asat{} \citep{singh2014astrosat} of four type 1 AGN: PG~0804+761 (hereafter PG0804), NGC~7469, SWIFT~J1921.1-5842 (hereafter SWIFT1921), and SWIFT~J1835.0+3240 (hereafter SWIFT1835). \asat{} is India's first space observatory that covers UV to X-rays with its suit of four co-aligned payloads: the Ultraviolet Imaging Telescope (UVIT; \citealt{tandon2017orbit,tandon2020additional}), the Soft X-ray Telescope (SXT; \citealt{singh2016orbit,singh2017soft}),  the Large Area X-ray Proportional Counters (LAXPC;  \citealt{yadav2016large, antia2017calibration})  and the Cadmium-Zinc-Telluride Imager (CZTI; \citealt{czti}). In this paper, we used the far UV and X-ray data simultaneously acquired with the UVIT and SXT, respectively. 

\begin{deluxetable*}{lccccc}
\tablenum{1}
\tablecaption{List of \asat{}/UVIT and SXT observations. The last column is the background-corrected net count rate of the sources in the $-2$ order of FUV gratings or $-1$ order of NUV grating. \label{tab:log}}
\tablewidth{0pt}
\tablehead{
\colhead{Source} & \colhead{ Observation } & \colhead{Instrument} &\colhead{Date of} & \colhead{Exposure time} & \colhead{Count rate}  \\
\colhead{name} & \colhead{ID} & \colhead{} &\colhead{observation} & \colhead{(ks)} & \colhead{(counts~s$^{-1}$)}
} 

\startdata
%MR~2251-178    &A04\_218T03\_9000002214  &UVIT/FUV-G1 &2018-07-09 \iffalse17:26:09.18 \fi  & 5.2 & $6.73\pm 0.04$\\
 %&A04\_218T03\_9000002214& SXT &2018-07-(09-10)&12.2 & $0.673\pm 0.008$ \\
% &&  && & \\
PG0804 &G07\_062T01\_9000001560 & \astrosat{}/UVIT/FUV-G2 & 2017-09-25 &4.1 &$8.9\pm 0.05$\\ 
& G07\_062T01\_9000001560& \astrosat{}/UVIT/NUV-Grating  & 2017-09-25 & 4.0 & $59\pm 0.1$\\
& G07\_062T01\_9000001560 & \astrosat{}/SXT & 2017-09-(25-26) & 14.2 & $0.208\pm0.004$ \\
%%&0605110101 & \xmm{}/EPIC-pn &2010-03-10 & 16.2 & $5.68\pm 0.02$\\
&&  && & \\
NGC~7469 & G08\_071T02\_9000001620 & \astrosat{}/UVIT/FUV-G1  &2017-10-18 \iffalse16:25:49.84 \fi & 3.4& $5.73\pm 0.04$  \\
\multicolumn{2}{c}{} & \astrosat{}/UVIT/FUV-G2  &  2017-10-18 \iffalse11:56:11.81 \fi &4.0 & $7.88\pm 0.05$    \\
&G08\_071T02\_9000001620& \astrosat{}/SXT &2017-10-(15-19)& 108 & $0.599\pm0.003$\\
&&  && & \\
SWIFT1921 &A04\_218T08\_9000002236 & \astrosat{}/UVIT/FUV-G1 & 2018-07-17 \iffalse20:35:41.05 \fi & 5.7 & $8.50\pm 0.04$\\
\multicolumn{2}{c}{} & \astrosat{}/UVIT/FUV-G2  & 2018-07-18 \iffalse09:26:31.6 \fi &5.4 & $9.72\pm 0.04$  \\
&A04\_218T08\_9000002236 & \astrosat{}/SXT &2018-07-(17-19)& 29 & $0.653\pm 0.005$ \\
&&  && & \\
SWIFT1835 &  A04\_218T04\_9000002086 & \astrosat{}/UVIT/FUV-G1  & 2018-05-10 \iffalse00:51:16.4 \fi &3.2 & $1.04\pm 0.02$ \\
&A04\_218T04\_9000002086 & \astrosat{}/SXT &2018-05-(9-10)& 20 & $0.338\pm 0.004$ \\
\enddata 
\end{deluxetable*}

\subsection{Ultra-Violet Imaging Telescope}
\label{subsec:uvitdata}
The  UVIT consists of two telescopes: one observes in the far ultra-violet band ($1200-1800$ \ang{}), referred to as the FUV channel. The other telescope observes in the near ultra-violet band ($2000-3000$ \ang{}) and the visible band ($3200-5500$ \ang{}), referred to as the NUV and VIS channels, respectively. The visible band is used to correct telescope drift while observing any source. Both the FUV and NUV channels have several broadband filters. In addition, the FUV channel contains two slitless low-resolution gratings (hereafter, FUV-G1 and FUV-G2) that are orthogonally oriented to each other. The NUV channel has only one slitless grating (hereafter, NUV-G). The spatial resolution of FUV and NUV broadband filter is $1-1.5^{\prime\prime}$. The full-width half maxima (FWHM) for the FUV gratings in the $-2$ order is  $\sim 14.3$ \ang{}, and that for the NUV grating in the $-1$ order is $\sim 33$ \ang{}. 

We described the UVIT data reduction in detail in the paper~I. Here, we briefly mention the steps. We obtained the level1 data from the \asat{} data archive\footnote{\url{https://astrobrowse.issdc.gov.in/astro_archive/archive/Home.jsp}} and processed using the CCDLAB pipeline software \citep{Postma_2017}. We extracted the source spectra from the $-2$ order of the FUV grating and $-1$ order of the NUV grating and the corresponding background spectra from a source-free region following the method described in \cite{Dewangan_2021}, and the tools available in the UVITTools.jl package\footnote{\url{https://github.com/gulabd/UVITTools.jl}}. We also used the response files made available as part of the UVITTools.jl package.

\subsection{Soft X-ray Telescope}
\label{subsec:sxtdata}
The SXT is a focusing X-ray telescope that uses conical mirrors to focus the X-ray photons onto a CCD detector \citep{singh2017soft}. It observes in the photon counting mode and is sensitive to the $0.3-7$~keV energy band. The field of view is $\sim 40^\prime$, and the energy resolution is $\sim 150\ev$  at $6\kev$.

We processed the level1 data using the SXT pipeline software AS1SXTLevel2-1.4b available at the SXT payload operation center (POC\footnote{\url{https://www.tifr.res.in/~astrosat\_sxt/sxtpipeline.html}}). This generates the clean event file for each orbit. We merged the clean event files using the SXT merger tool  SXTMerger\footnote{\url{https://github.com/gulabd/SXTMerger.jl}}. We extracted the source spectra from the final clean image file using the tool XSELECT available in the HEASoft package (version 6.29).  We used the background spectrum (SkyBkg\_comb\_EL3p5\_Cl\_Rd16p0\_v01.pha), instrument response (RMF: sxt\_pc\_mat\_g0to12.rmf), and effective area (ARF: sxt\_pc\_excl00v04\_20190608.arf) from the SXT POC website. We grouped each  PHA spectral dataset with a minimum $25~\rm counts~bin^{-1}$ using the ftool FTGROUPPHA available within HEASoft. 

\begin{figure*}
\centering
  \includegraphics[scale=0.5]{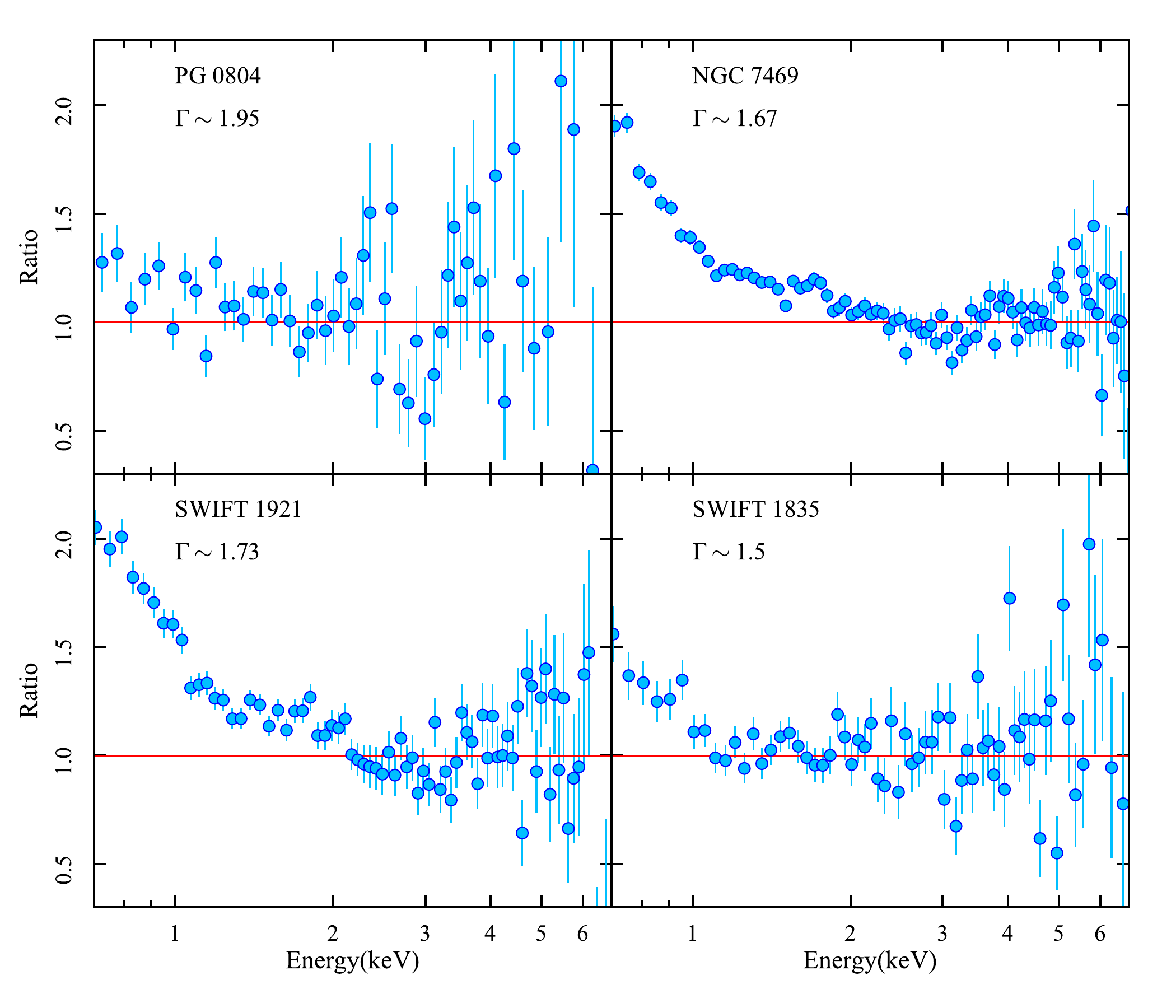}
\caption{Soft X-ray excess observed in the four objects shown by the plotting ratio (data/model). The power law (modified by Galactic absorption) is fitted between the $2-10$ keV band and then extrapolated to 0.7 keV. The respective $\Gamma$ are 1.95 (PG0804), 1.67 (NGC 7469), 1.73 (SWIFT1921), and 1.5 (SWIFT1835).  }
\label{fig:softEx}
\end{figure*}

%%\subsection{XMM-Newton}
%%\label{subsec:xmmdata}

%%We obtained the observation data files (ODF) generated from the \xmm{} observations of PG~0804+761 
%We obtained the EPIC-pn (European Photon Imaging Camera) observation data files (ODF) on PG~0804+761 
%%from the  \xmm{} data archive\footnote{\url{https://nxsa.esac.esa.int/nxsa-web/\#search}}. We processed the EPIC-pn data using the \xmm{} Science Analysis System (SAS; \citealt{gabriel2004xmm}) version 21.0.0. We generated a clean event file using the task epproc. We extracted a light curve in the $10-12$ keV energy band from the full field and examined the presence of particle-induced flares.  We removed the particle flaring intervals where the count rate exceeded 0.4 count~s$^{-1}$. This resulted in a final exposure time of 16.2~ks (see Table~\ref{tab:log}). We found that the data are not affected by the pile-up. We generated the source spectrum from a circular region of radius 40$^{\prime\prime}$ centered at the source position. We extracted the background spectrum from two source-free circular regions with a radius of 20$^{\prime\prime}$ each. We generated the response files using the rmfgen and arfgen tasks.  

\begin{deluxetable*}{clcccc}
\tablenum{2}
\tablecaption{Best-fit parameters of the \opt{} and emission line components fitted to the UV/X-ray spectra.}
\label{tab:optagn}
\tablewidth{700pt}
\tablehead{
\colhead{Model}&\colhead{Parameters}  &   \colhead{PG0804} & \colhead{NGC~7469} &   \colhead{SWIFT1921} & \colhead{SWIFT1835}
}
\startdata
%\multirowcell{3}{OPTXAGNF\\a=0.998}
\multirow{1}{*}{\begin{tabular}{c}\opt 
                  \end{tabular}}
&$\log (L/L_{Edd})$  & $-0.270_{-0.003}^{+0.003}$ & $-0.18_{-0.09}^{+0.20}$ & $-0.34_{-0.04}^{+0.06}$ & $-1.8_{-0.1}^{+0.1}$\\
& $a^\star$ &  $0.998$ (f) & $0.37_{-0.30}^{+0.29}$ & $0.998$ (f) & 0.998 (f)\\
& $r_{cor}$  & $1.6_{-0.1}^{+0.1}$ & $39_{-19}^{+20}$ & $95.3_{-10.2}^{+11.7}$ & $16.3_{-5.8}^{+12.2}$ \\
& $kT_w$  & $0.26$ (f) & $0.31_{-0.04}^{+0.05}$  & $0.12_{-0.02}^{+0.02}$ & $0.12_{-0.04}^{+0.06}$\\
& $\tau$  & $8$ (f) & $13.3_{-3.1}^{+2.1}$ & $26.5_{-3.9}^{+5.6}$ & $>20$\\
& $\Gamma$  & $2.0_{-0.1}^{+0.1}$ & $1.78_{-0.17}^{+0.16}$ & $2.04_{-0.05}^{+0.05}$ & $1.52_{-0.07}^{+0.07}$\\
& $f_{pl}$  & $>0.3$ & $0.17_{-0.07}^{+0.23}$ & $0.35_{-0.04}^{+0.03}$ & $0.8_{-0.2}^{+0.2}$  \\
\multirow{1}{*}{\begin{tabular}{c}\xab 
                  \end{tabular}}
& $N_H$($10^{22}{\rm~cm^{-2}}$) & -- & $3.3_{-0.6}^{+0.3}$ & 36 (f) & -- \\
& $\log \xi$ &   -- & $<-1.6$  & 0.2 (f)& --\\
& $f_{c}^{xabs}$    & -- & $0.87_{-0.07}^{+0.04}$  & 0.47 (f)& --\\
Fe K$\alpha$ & norm ($10^{-4}$) & -- & $0.7_{-0.4}^{+0.4}$ & -- & --\\
$\chi^2/dof$ & & 454/391 & 395.9/361 & 362.9/324& 262.0/238\\
\enddata
%\tablecomments{ }
\end{deluxetable*}
\begin{deluxetable*}{clccccccc}
\tablenum{3}
\tablecaption{Best-fit parameters of \fagn{} model fitted to \asat{} spectral data. The inclination angles are fixed at $30^\circ$ (PG0804), $20^\circ$ (NGC~7469),  $ 31^\circ$ (SWIFT1921), and $ 30^\circ$ (SWIFT1835). The maximum height of the corona is fixed at 10 $r_g$, and the hot corona temperature at 100~keV.}
\label{tab:fagnsed}
\tablewidth{700pt}
\tablehead{
\colhead{FAGNSED}&$\log\frac{\dot{M}}{\dot{M_{Edd}}}$  & $a^\star$ & $kT_{warm}$ &  $\Gamma_{hot}$& $\Gamma_{warm}$ & $R_{hot}$ & $R_{warm}$& $\chi^2/dof$
}
\startdata
PG0804 & $-0.86_{-0.07}^{+0.05}$ & $^{\star}0.76_{-0.20}^{+0.08}$ & $0.26$ (f) & $1.94_{-0.15}^{+0.15}$ & $3.3_{-0.2}^{+0.3}$& $^{\star}5.4_{-0.9}^{+1.9}$& $55.6_{-15.2}^{+48.6}$ & 436.4/390 \\
NGC 7469 & $-0.39_{-0.08}^{+0.08}$ & $<0.67$ & $0.27_{-0.03}^{+0.04}$ & $1.89_{-0.18}^{+0.16}$ & $2.28_{-0.13}^{+0.19}$ & $11.7_{-5.4}^{+3.8}$ & $252_{-106}^{+90}$ & 391.6/361\\
SWIFT1921& $-0.19_{-0.03}^{+0.21}$&$<0.86$ & $0.12_{-0.02}^{+0.04}$ & $2.03_{-0.05}^{+0.05}$ & $2.7_{-0.2}^{+1.2}$ & $14.1_{-7.1}^{+1.0}$ & $330_{-274}^{+42}$& 364.0/323\\
SWIFT1835 & $-1.8_{-0.1}^{+0.1}$ & 0.998 (f) & $0.12_{-0.02}^{+0.07}$ & $1.52_{-0.08}^{+0.05}$ & $<2.25$ & $9.9_{-1.9}^{+3.7}$ & $18.9_{-3.9}^{+7.2}$& 259.6/238\\
\enddata
\tablecomments{ $^\star$ 68\% confidence interval.}
\end{deluxetable*}

\begin{figure*}
\centering
\gridline{\fig{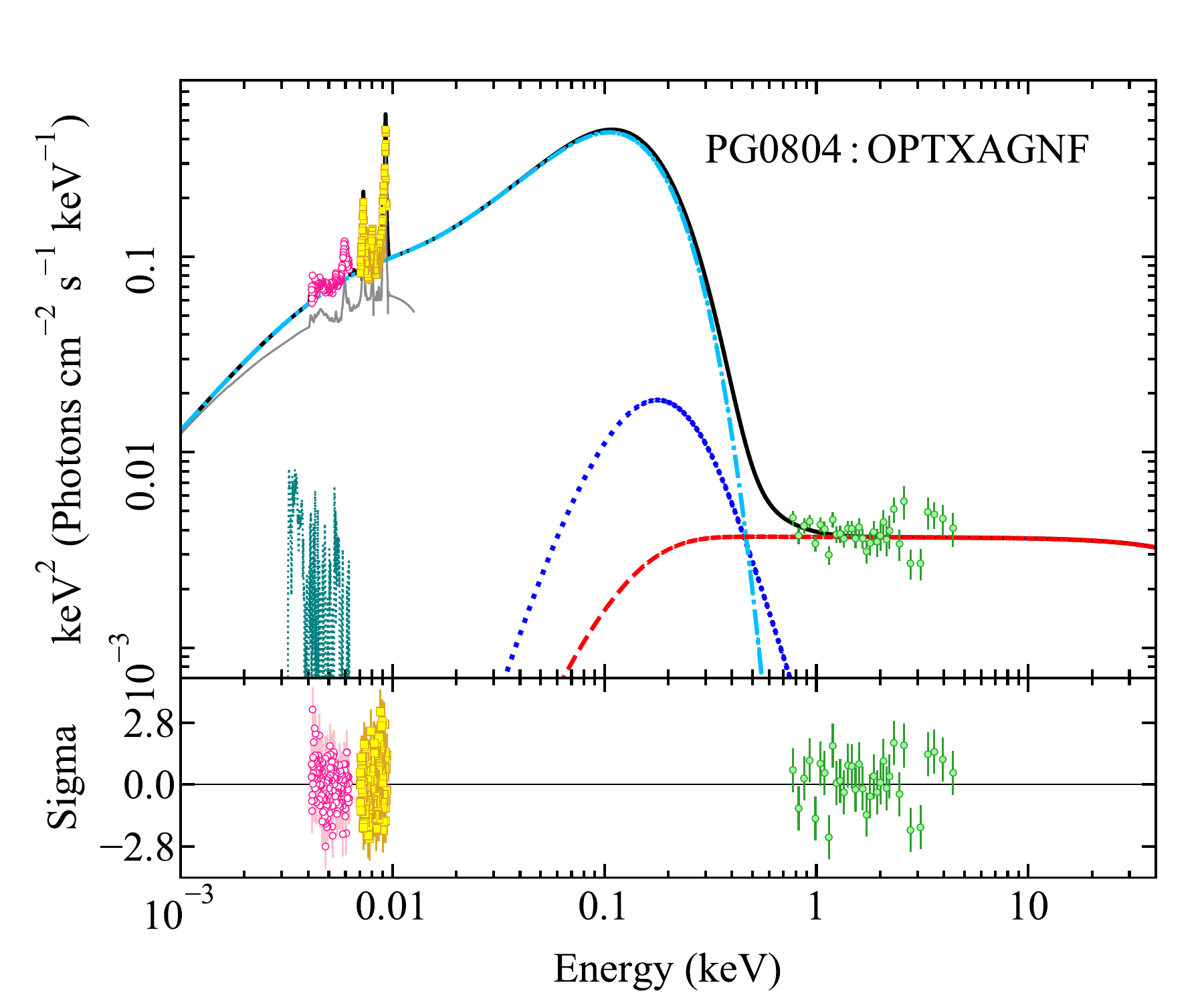}{0.53\textwidth}{(a)}
           \fig{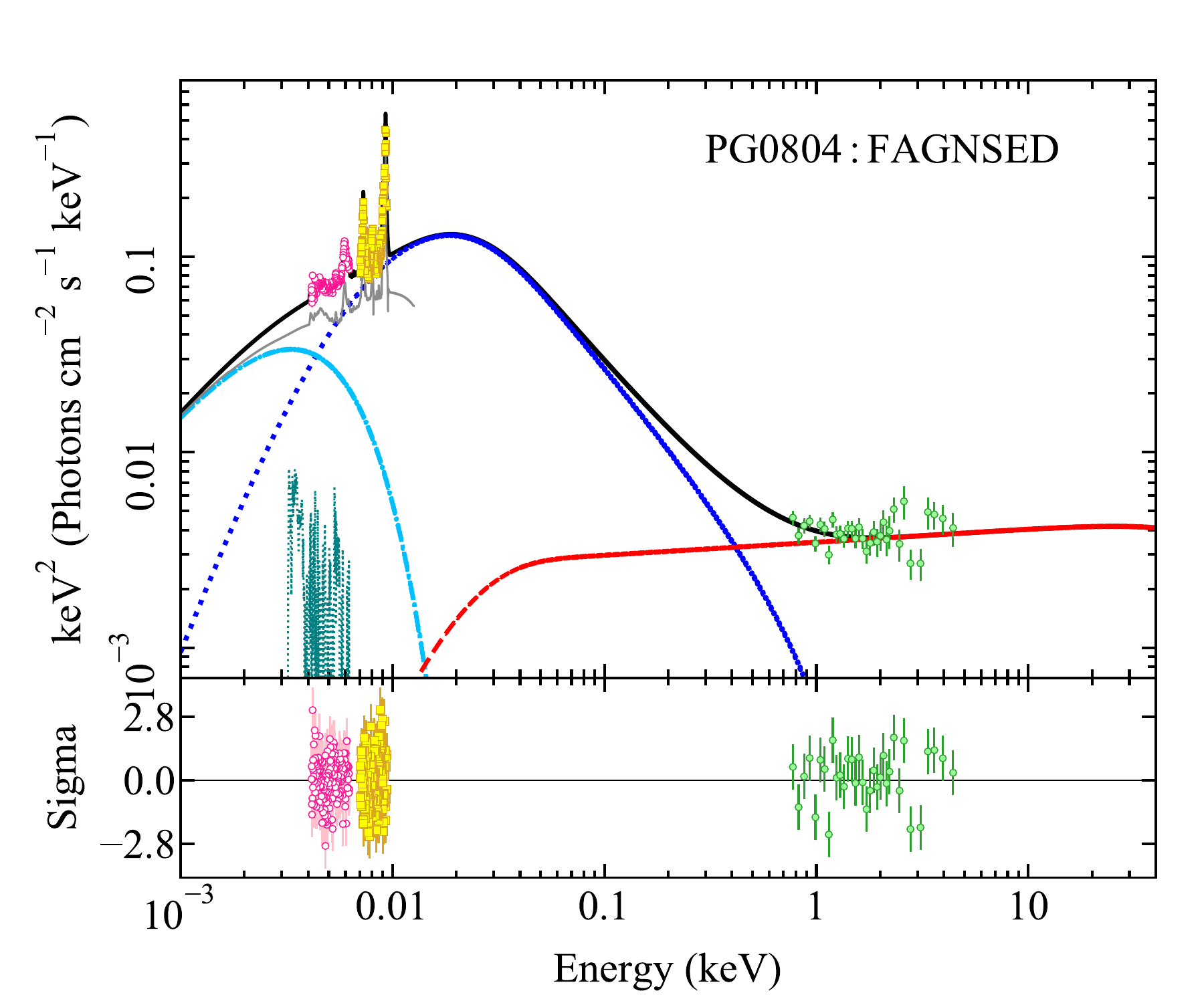}{0.53\textwidth}{(b)}
          }
\caption{Results of UV/X-ray broadband spectral analysis. \textbf{Upper panels:} The best-fit total unabsorbed model (black), absorbed model (gray), and the absorption-corrected spectral datasets UVIT/FUV grating (yellow), UVIT/NUV grating (pink), and the SXT (green). Also shown are the different model components: accretion disk  (cyan), soft excess  (blue), X-ray power law (red), and the \ion{Fe}{2} complex (teal). \textbf{Lower panels:} The fit residuals in terms of $data-model/error$.}
\label{fig:joint_eem_pg08}
\end{figure*}

\section{UV -- X-ray joint spectral analysis} \label{sec:spec_analys}

In paper~I, we analyzed the UVIT grating spectra of the AGN listed in Table~\ref{tab:log}. We accounted for the intrinsic and Galactic extinction, host galaxy contribution, BLR/NLR emission, \ion{Fe}{2} emission, and obtained the intrinsic UV continuum emission. We fitted the continuum with a simple multi-temperature disk blackbody model \dbb, which allowed us to estimate the peak inner disk temperatures. Then, we replaced \dbb{} with \opt{} (disk component only) to infer the inner disk geometry. 

In this paper, we use the best-fit model with the continuum component as \opt{} and construct the broadband SED by fitting the SXT and UVIT spectral data jointly. 
%The redshifts of the sources are 0.016 (NGC~7469), 0.037 (SWIFT1921), 0.1 (PG0804), and 0.058 (SWIFT1835).
%The black hole masses adopted for these four objects are  $10^7$\msun (NGC~7469; \citealt{2004ApJ...613..682P}), $5.4\times 10^8$\msun (PG0804; \citealt{2015PASP..127...67B}), $1.25 \times 10^7$\msun (SWIFT1921; \citealt{Wang_2007}) and $10^9$\msun (SWIFT1835; \citealt{10.1111/j.1365-2966.2004.07822.x}).

We initially analyze the X-ray spectrum for each AGN to investigate the presence of different spectral components, such as the soft X-ray excess, warm, or neutral absorbers. Next, we model the UVIT/SXT spectra jointly. In this case, we fix the parameters associated with the emission and absorption lines, as well as warm and neutral absorbers, to those obtained during separate UV and X-ray spectral fittings. We use \opt{} \citep{done2012intrinsic} or \fagn{}\footnote{\url{https://github.com/scotthgn/fAGNSED}} \citep{kubota2018physical, hagen2023estimating}  models to represent the underlying UV/X-ray continuum. \fagn{} model is an upgraded version of \opt{} model. In \opt{}, the standard accretion disk is truncated at a radius $r_{cor}$, below which the disk seed photons are Comptonized in a warm ($kT_{w} \sim 0.1-1$ keV) and hot ($kT = 100$~keV) corona to produce the soft X-ray excess and the X-ray power-law components, respectively. The disk emission beyond the $r_{cor}$ emits as a modified blackbody emission. However, unlike \opt{}, the Comptonization region is radially stratified into warm and hot corona in \fagn{}. The warm Comptonizing corona exists between $R_{warm}$ and $R_{hot}$ over a passive disk, and the inner hot flow extends from $R_{hot}$ to $R_{ISCO}$. The sum total of Comptonized disk photons in each radial bin produces the overall soft excess emission. Therefore, \fagn{} provides the radial extent of each emitting region and improves our understanding of the accretion disk better than the \opt{} model. The free parameters of \opt{} model are: logarithm of the Eddington ratio ($\log(L/L_{Edd})$), black hole spin ($a^\star$), coronal radius ($r_{cor}$), warm corona temperature ($kT_{w}$), optical depth of the warm corona ($\tau_w$),  X-ray photon index ($\Gamma$), and the fraction of power law emission below $r_{cor}$ ($f_{pl}$). 
%Since we have broad spectral coverage with FUV and NUV grating spectra for PG0804, 
The relevant parameters of \fagn{} are the logarithm of the Eddington ratio, spin parameter ($a^\star$), the inclination angle, the temperature of the warm corona ($kT_w$), photon index of the X-ray power-law ($\Gamma_{hot}$) and soft X-ray excess ($\Gamma_{warm}$), outer radius of the hot corona ($R_{hot}$) and the radius of warm corona ($R_{warm}$), and the maximum height of the X-ray corona. We fixed the normalization to 1 for both these models and varied the rest of the parameters during the joint UV--X-ray spectral fitting unless mentioned otherwise. The errors are quoted at a 90\% confidence interval unless mentioned otherwise.

\subsection{PG~0804$+$761}

We fitted the $2-7$~keV SXT spectral data with the Galactic-absorbed power-law model. Extending the data and model to 0.7 keV, we observed some soft X-ray excess emission (see Fig.~\ref{fig:softEx}). We added a \zbb{} model in the $0.7-7$~keV band to account for this excess emission.  This did not improve the fit significantly ($\Delta \chi^2 = 2$), possibly due to the low signal-to-noise of the SXT data. With only the Galactic absorbed power-law in the $0.7-7$~keV band, the final best fit $\chi^2$ per degree of freedom ($dof$) $= 60/52$.

\begin{figure*}
\centering
\gridline{\fig{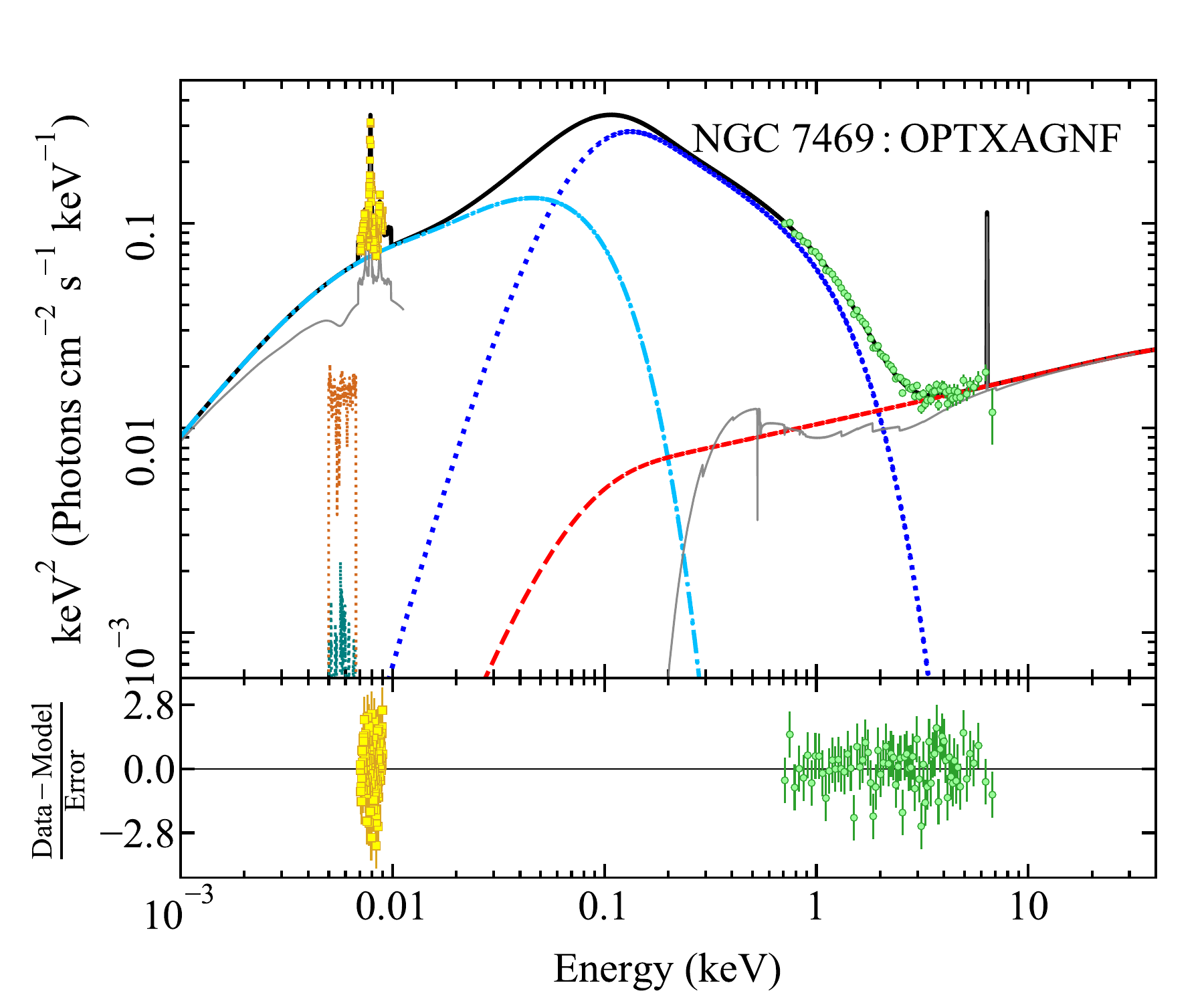}{0.53\textwidth}{(a)}
        \fig{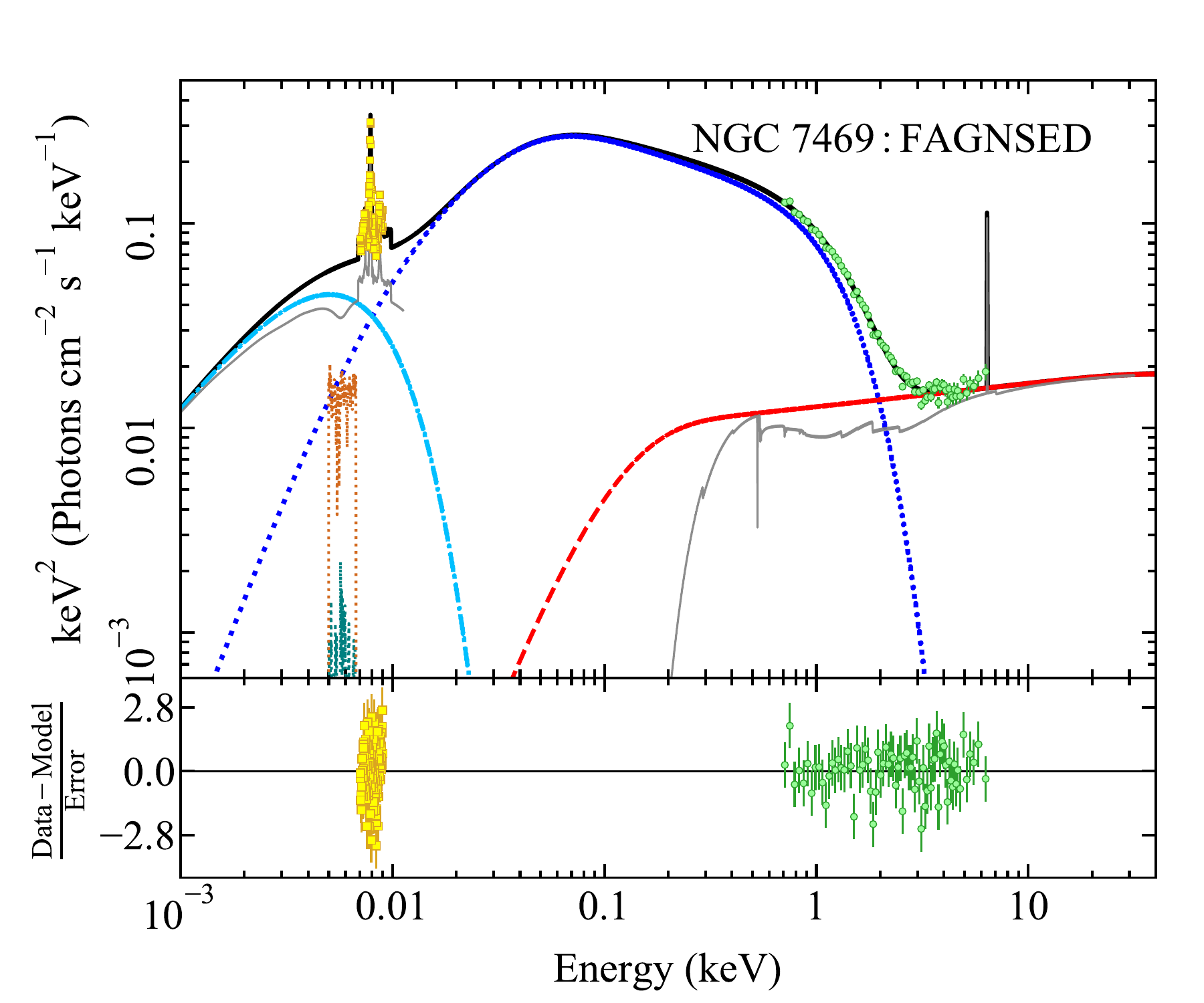}{0.53\textwidth}{(b)}}
\caption{Same as Fig.~\ref{fig:joint_eem_pg08} but for NGC 7469. The additional model component, star-burst SB3 template, is shown in orange.}
\label{fig:joint_eem_n7469}
\end{figure*}

Next, we included previously fitted (with \opt{}) FUV-G1 and NUV-G spectra to the SXT spectral data. We removed the \texttt{ZPOWERLAW} and the \texttt{ZBBODY}, which is accounted for by the \opt{} model component. The model expression for the joint UV/X-ray spectral fitting in XSPEC is \texttt{CONSTANT $\times$ TBABS $\times$ REDDEN $\times$ GABS $\times$ [PLABS(OPTXAGNF) + GAUSSIAN$_{UV}$]}. Since the soft excess is weak in the SXT spectrum, we fixed the $kT_w$ at 0.26 keV and the $\tau$ at 8 following \citet{2018A&A...611A..59P}. We obtained a lower limit in the $f_{pl}>0.3$.
%%Therefore, we fixed the $kT_{w}$ at 0.262~keV during the error calculation. We obtained a lower limit of 0.992 for the $a^\star$. Therefore, we fixed the $a^\star$  at 0.998. 
We obtained the final best fit $\chi^2/dof = 454/391$ with \opt{} (see Table \ref{tab:optagn}). The unabsorbed data, SED, and absorbed SED are shown in Fig.~\ref{fig:joint_eem_pg08}(a).

We also modeled the UV/X-ray  spectral data with the \fagn{} model. We replaced only the \opt{} model component with the \fagn{}. 
%As before, we used the \texttt{PLABS} component for the SXT and EPIC-pn data to account for the difference in the spectral index. 
The inclination angle is fixed at $30^\circ$, as we obtained an upper limit of $60^\circ$. We found the $\chi^2/dof = 436.4/390$ with \fagn{} model. 
We found the spin parameter, $a^\star = 0.76_{-0.20}^{+0.08}$ (1$\sigma$ error), and the $\rm R_{hot} = 5.4_{-0.9}^{+1.9}~r_g$ (1$\sigma$ error).
%We found the $R_{hot}<10 r_g$; therefore, we tied the height of the corona to $R_{hot}$. 
The remaining best-fit parameters of \fagn{} are listed in Table~\ref{tab:fagnsed}. In Fig.~\ref{fig:joint_eem_pg08}(b), we show the unabsorbed and absorbed SED and the model components. 

%\end{comment}

\subsection{NGC~7469}

\begin{figure*}
\centering
\gridline{\fig{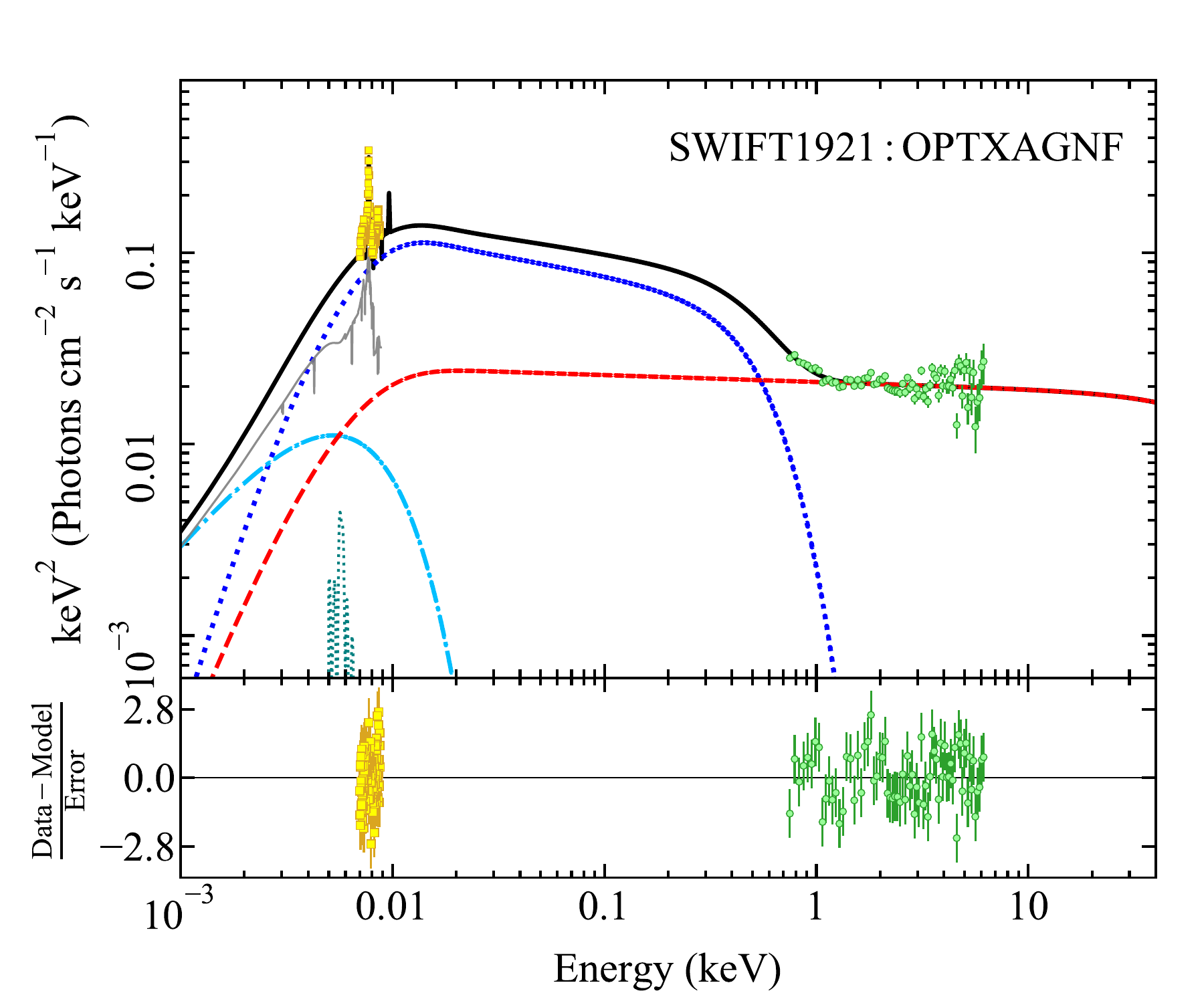}{0.53\textwidth}{(a)}
           \fig{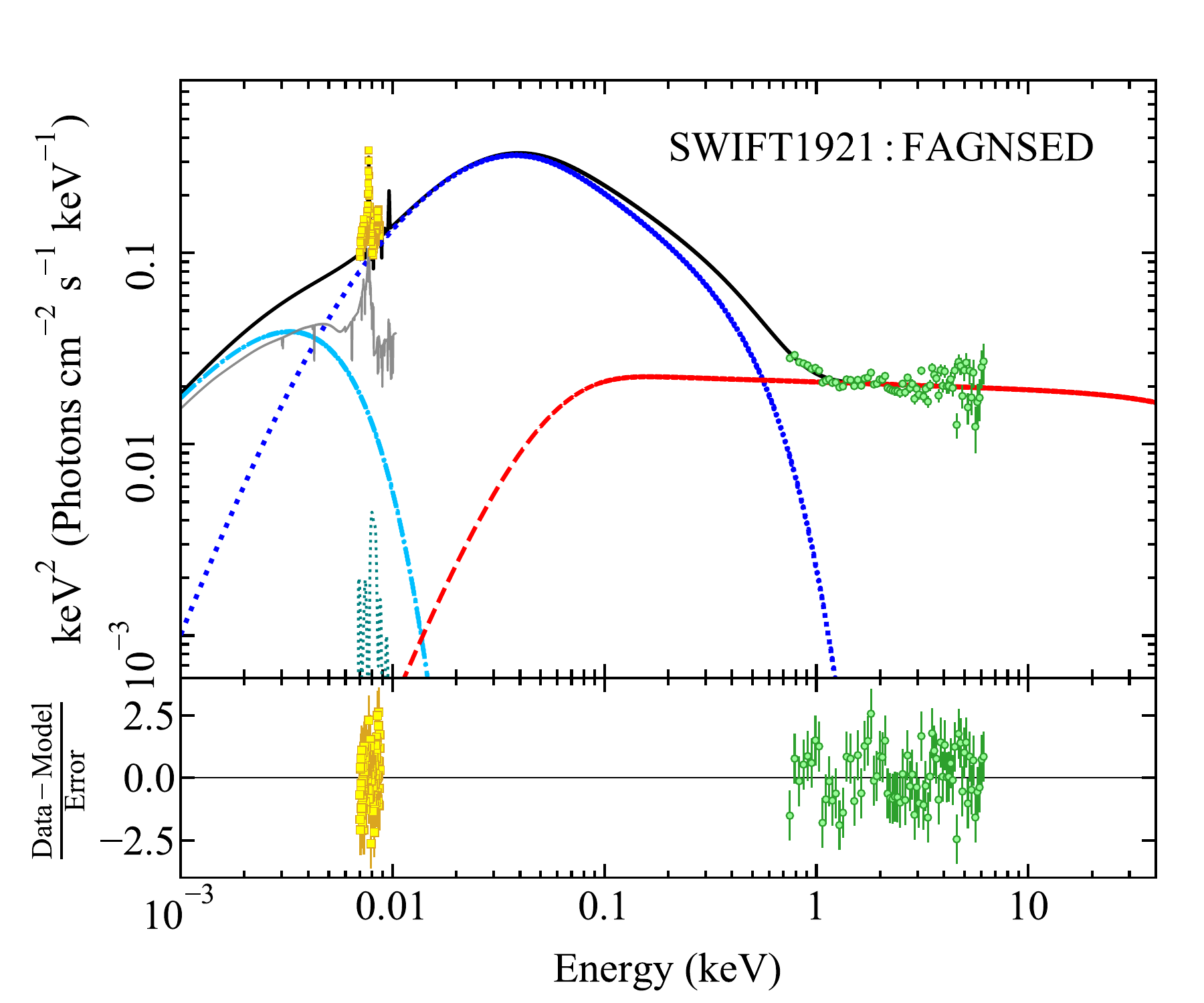}{0.53\textwidth}{(b)}
          }
\caption{Same as Fig. \ref{fig:joint_eem_pg08} but for SWIFT1921.}
\label{fig:joint_eem_s1921}
\end{figure*}

We modeled the X-ray spectrum in the $2-7$ keV range with Galactic absorbed power-law and a narrow ($\sigma = 10$~eV) Fe~K$\alpha$ emission line at $\sim 6.5\kev$. The soft X-ray excess emission is apparent above the power-law ($\Gamma \sim 1.67$; Fig.~\ref{fig:softEx}) at low energies ($\sim 2$~keV). Adding a \zbb{} component to account for the soft excess emission improved the $\chi^2$ by 94 (in $0.7-7$~keV band) for two additional free parameters, blackbody temperature (kT) and the normalization. 
The \xmm-RGS and \chandra-HETGS spectra of NGC~7469 showed the presence of multi-layer warm absorber components with column density and the ionization varying in the range $N_H \sim 0.7-5.2 \times 10^{21}$, and  $\log \xi \sim 1.9-3.3$, respectively \citep{mehdipour2018multi,grafton2020multi}. Further, they found the total column density considering all the absorbers to be similar over time, although the ionization levels of the different components varied slightly. Therefore, we also tested the presence of a warm absorber using the \xab{} model by varying the $N_H$ within the range provided by \citet{mehdipour2018multi}. Since we could not constrain the column density, we fixed the $N_H$ to the highest value observed by \citet{grafton2020multi}, $5 \times 10^{21}$ \cmsq{}, turbulent velocity $v$ at $100$\kms{} and the redshift at 0.016 \citep{grafton2020multi}. This resulted in a marginal improvement in the statistic, $\Delta \chi^2 = 5$ for one additional free parameter, $\log \xi = 3.21_{-0.34}^{+0.49}$. Further addition of \xab{} component did not change the statistics. Therefore, we included only one warm absorber component. The final XSPEC model expression for the SXT spectrum is \texttt{TBABS $\times$ XABS $\times$[ZPOWERLAW + FeK$\alpha$ + ZBBODY]}. We obtained the final $\chi^2/dof = 129.6/84$ with a gain shift of 61~eV and a systematic error of $2\%$. 

Next, we included the previously fitted UVIT grating spectra with \opt, emission/absorption lines (\texttt{GAUSSIAN$_{UV}$/GABS}), \ion{Fe}{2} emission, and star-burst emission (\texttt{SB3}), and removed the \zpl{} and \zbb{}. All these components are corrected for Galactic extinction. Again, we used $2\%$ systematic error and a gain shift fixed at that obtained during the SXT spectral analysis. 
%The spin parameter is fixed to 0, as we obtained an upper limit of 0.5. 
%We also fixed the $\Gamma$ to that obtained in the $2-7$~keV band data ($\Gamma = 1.67_{-0.05}^{+0.05}$) with Galactic absorbed power-law, as the broadband data resulted in an unphysically small lower bound of 1.16 ($\Gamma = 1.42_{-0.26}^{+0.22}$). 
We obtained an unusually flat spectrum ($\Gamma \sim 1.4$). Therefore, we varied the $N_H$ in the \xab{} model, which was fixed at $5 \times 10^{21}$ \cmsq{}. We also varied the ionization parameter $\log{\xi}$ and the covering fraction $f^{xabs}_c$. We found the best-fit value for the $N_H \sim 3.3 \times 10^{22}$ \cmsq{} with the absorber being neutral.  This $N_H$ is slightly larger than that obtained previously \citep{mehdipour2018multi,peretz2018multi,grafton2020multi}. This could have resulted due to the poor spectral resolution of SXT. However, we found the best-fit photon index $\sim 1.78$ and the spin parameter $\sim 0.37$ (Table \ref{tab:optagn}).   
%We can only obtain the lower limit of 0.36~keV for the temperature of the warm plasma. This is consistent with that obtained by \citet{grafton2020multi}, who found the $kT \sim 0.6$~keV in their observations. 
We obtained the final  $\chi^2/dof = 395.9/361$ with the model expression  \texttt{REDDEN $\times$ TBABS $\times$ [SB3 + \texttt{XABS  $\times$ GABS $\times$ (OPTXAGNF + GAUSSIAN$_{UV}$ + FeK$\alpha$)]}}. 
The unabsorbed data, SED, and fit residuals are shown in Fig.~\ref{fig:joint_eem_n7469}(a). With the \fagn{} model as a UV -- X-ray continuum component, we found a substantially large emitting region contributing to the soft excess (see Fig~\ref{fig:joint_eem_n7469}(b)). We obtained the best-fit $\chi^2/dof = 391.6/361$ with the \fagn{} model (Table~\ref{tab:fagnsed}). We showed the unabsorbed data, model, and fit residuals in Fig.~\ref{fig:joint_eem_n7469}(b).

\subsection{SWIFT~J1921.1$-$5842}

\begin{figure*}
\centering
\gridline{\fig{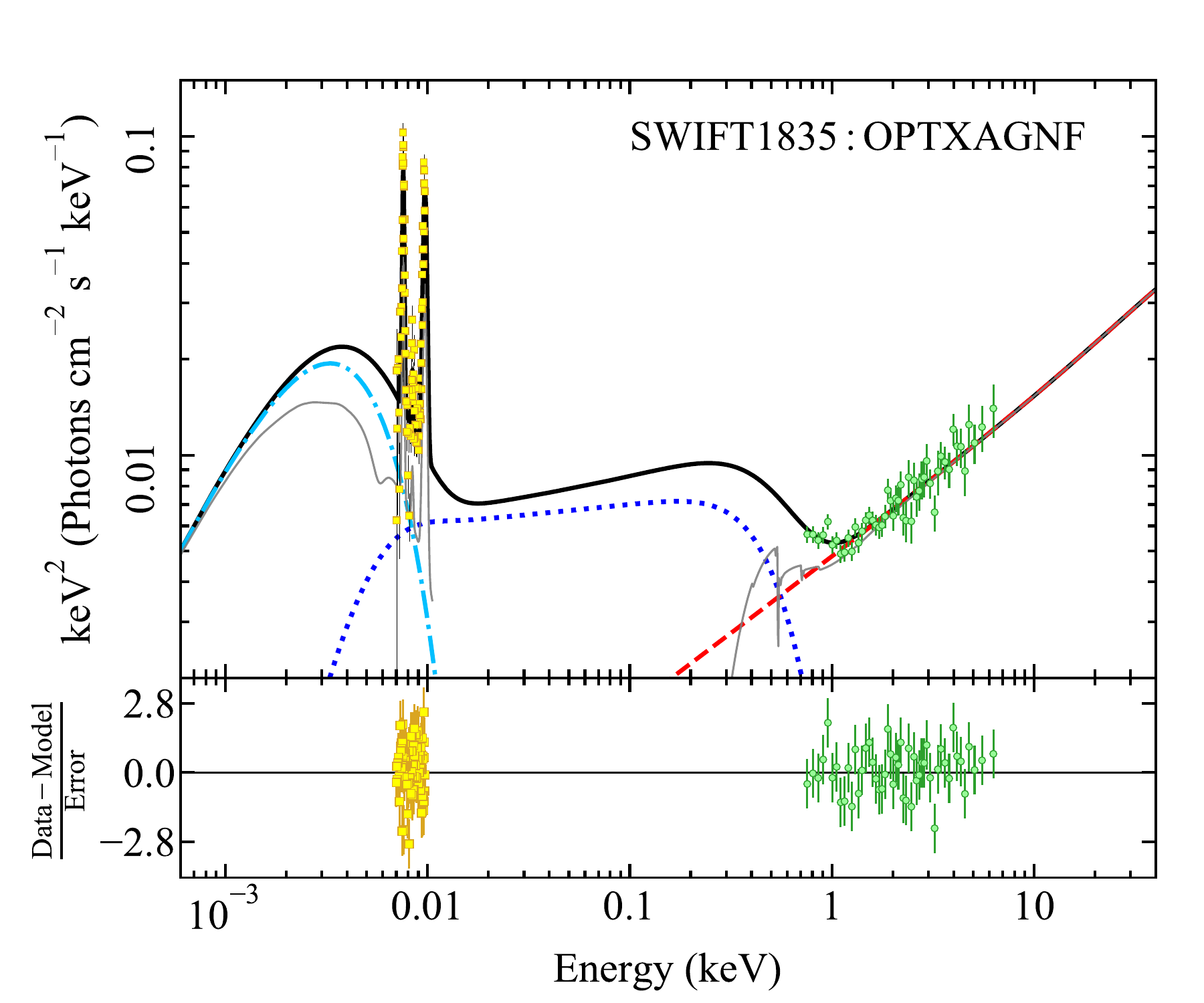}{0.53\textwidth}{(a)}
        \fig{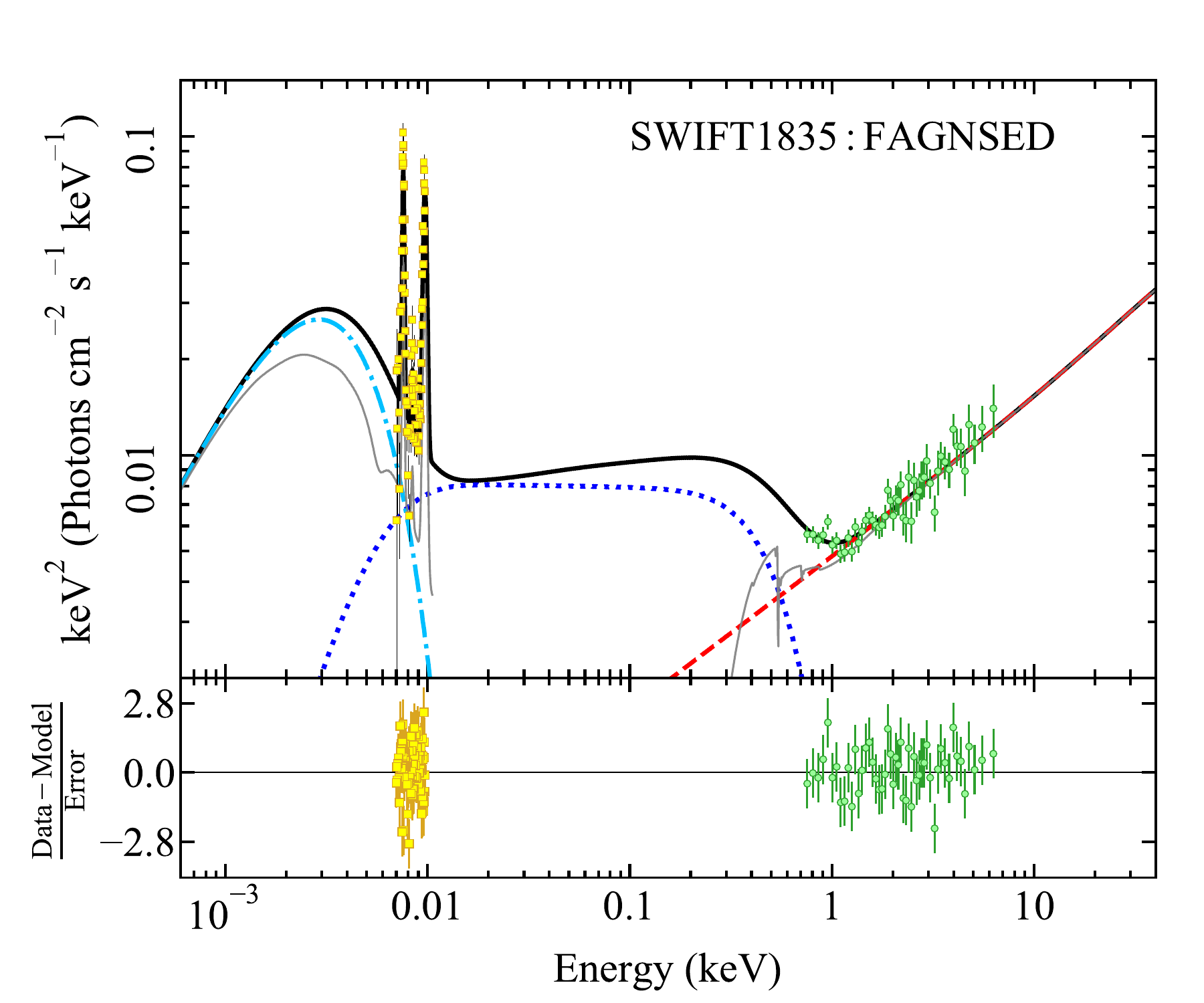}{0.53\textwidth}{(b)}}
\caption{Same as Fig. \ref{fig:joint_eem_pg08} but for SWIFT1835.}
\label{fig:joint_eem_s1835}
\end{figure*}

We fitted the SXT spectrum with a Galactic absorbed power-law in the $2-7$ keV band. We observed an excess emission over the power-law ($\Gamma \sim 1.7$; see Fig.~\ref{fig:softEx}) below 2~keV. We used a \zbb{} to account for this excess emission. This improved the $\chi^2$ by $57$ (in $0.7-7$~keV band) for two additional free parameters. Next, we incorporated the warm absorber model \xab{} to investigate the presence of this component. This improved the  $\chi^2$ by $9$ for three additional free parameters, absorption column density ($N_H$), covering fraction ($f_{c}^{xabs}$), and the ionization ($\log \xi$). We obtained an upper limit on the ionization parameter ($\log \xi< 2.1$) of the warm absorber. Therefore, we fixed the ionization parameter $\log \xi$ at 0.2 as obtained by \citet{ghosh2020broad} in their broadband SED modeling utilizing non-simultaneous \xmm{}  and \textit{NuStar} observations. We found the warm absorber column density $N_H = 3.6_{-0.1}^{+0.5} \times 10^{23} \rm cm^{-2}$ and $f_{c}^{xabs} = 0.5_{-0.1}^{+0.5}$.  We obtained the final $\chi^2/dof= 84/67$ after using a $\sim 40$~eV gain shift to the SXT spectral data using the \texttt{gain fit} command in XSPEC. The final model expression in XSPEC for the SXT spectrum: \texttt{TBABS $\times$ XABS $\times$ (ZPOWERLAW + ZBBODY)}.   

Next, we included the UVIT/grating spectra to construct the broadband SED. We used the best-fit model consisting of \opt{} as the continuum component from paper~I. The other model components include emission and absorption lines from the BLR/NLR. We fixed the UV emission and absorption line parameters and the cross-normalization constant between the gratings. Also, we fixed the \xab{} model parameters to those obtained during the SXT spectral fitting, as varying these parameters during the joint modeling have no effect on the statistic. We could not constrain the spin parameter, which we fixed to 0.998 \citep{ghosh2020broad}. The final model is \texttt{REDDEN $\times$ TBABS $\times$ XABS $\times$ (OPTXAGNF + GAUSSIAN$_{UV}$)}. We obtained $\chi^2/dof = 362.9/324$  in the joint UV -- X-ray spectral modeling (Table \ref{tab:optagn}). In Fig.~\ref{fig:joint_eem_s1921}(a), we showed the unabsorbed SED, data, and fit residuals. The \fagn{} model as a broadband continuum component resulted in a similar fit as with the \opt{} model. We listed the best-fit parameters in Table~\ref{tab:fagnsed}. The unabsorbed data, SED and fit residuals are shown in Fig.~\ref{fig:joint_eem_s1921}(b).

\begin{comment}
\begin{deluxetable*}{lccccccccc}
\tablenum{4}
\tablecaption{Model integrated unabsorbed continuum fluxes of different emission components based on the \opt{} model. All the fluxes are in units of \ergS.\label{tab:flux_comp}}
\tablewidth{0pt}
\tablehead{
Sources & Mass  & Distance & $F_{disk}$  & $F_{soft}$  &$F_{hard}$ & $\frac{L_{Bol}}{L_{Edd}}$& $\rm \frac{L_{disk} (0.001-0.1~keV)}{L_{bol}(0.001-100~keV)}$
\\
 & & &$0.001-0.01$ keV &$0.5-2$ keV & $2-10$ keV&
 \\
 & ($10^8$\msun)& (Mpc) & ($10^{-10}$) & ($10^{-11}$) &($10^{-11}$) & }
\startdata
%MR~2251   &  &  &   \\
PG0804 & 5.4$^a$ & 447.5&$1.85$ & $0.3$ & $1.2$ & 0.54& 0.63\\
NGC~7469 & 0.1$^b$ & 68.7 &$1.39$ & $1.39$  &$3.8$ & 0.72 & 0.32   \\%for Gamma 1.8
SWIFT1921 & 0.39$^c$ &158 &$0.29$ & $1.4$ &$5.2$ & 0.46& 0.03 \\
SWIFT1835 & 10$^d$& 233.8 & $1.08$ & $0.2$ & $2.8$ & 0.02& $0.15$\\
\enddata
\tablecomments{Black hole masses are taken from: a -- \citealt{2015PASP..127...67B}; b --\citealt{2004ApJ...613..682P}; c --\citealt{Wang_2007}; d --\citealt{10.1111/j.1365-2966.2004.07822.x}}
\end{deluxetable*}
\end{comment}

\subsection{SWIFT~J1835.0$+$3240}

 The soft excess emission is shown in Fig.~\ref{fig:softEx} over the Galactic-absorbed power-law in the energy range of $2-7$~keV.  Adding a \zbb{} improved the $\chi^2$ by 16 (in $0.7-7$~keV band) for two additional free parameters and the $\chi^2/dof = 79/65$. We also tested for the presence of a warm absorber component by fixing the parameters to the values obtained by \citet{ursini2018radio}. The addition of this component (\xab) significantly worsened the fit. Therefore, we varied the covering fraction of the absorber. This resulted in a covering fraction close to zero. We fixed the covering fraction to 1 and varied the ionization parameter. The $\chi^2$ remained the same as that without the warm absorber model, and we could obtain the lower limit of 3.32 for $\log \xi$.  Therefore, we did not include this component for this source. We obtained the final $\chi^2/dof = 79/65$ after using a gain shift of 34~eV for the SXT spectral data. The XSPEC model expression is \texttt{TBABS $\times$ (ZPOWERLAW + ZBBODY)}. 

 Next, we added the UVIT/grating (FUV-G2) spectrum with the best-fit model components being \opt{} and three emission lines corrected for Galactic reddening. We obtained similar statistics for the spin parameter at 0.998 and 0, although the Eddington ratio and the $r_{cor}$ differ in each case. 
  
 We obtained the  $L/L_{Edd}\sim 0.02$ and $r_{cor} \sim 51 ~r_g$ for a$^\star = 0$, and, $L/L_{Edd} \sim 0.01$ and $r_{cor} \sim 14~ r_g$ for a$^\star = 0.998$. For both the spin cases, the photon index $\Gamma~ (\sim 1.56)$ and $\chi^2/dof~ (= 262/238)$ remained unchanged. The model expression with \opt{} as the continuum component is \texttt{REDDEN $\times$ TBABS $\times$ (OPTXAGNF + GAUSSIAN$_{UV}$}). In Table~\ref{tab:optagn} we listed the best fit parameters with the spin parameter fixed at 0.998. The unabsorbed SED, data, and the residuals are shown in Fig.~\ref{fig:joint_eem_s1835}(a). With \fagn{} as a UV-X-ray continuum component, we found similar best-fit model parameters (see Table~\ref{tab:fagnsed}). We showed the unabsorbed data, SED, and residuals in Fig.~\ref{fig:joint_eem_s1835}(b).

\begin{comment}
\begin{figure*}
\centering
\gridline{\fig{jointmr22.pdf}{0.33\textwidth}{(a)}
        \fig{jointn7469.pdf}{0.33\textwidth}{(b)}
           \fig{jointpg08.pdf}{0.33\textwidth}{(c)}
          }
\gridline{\fig{joints1921.pdf}{0.33\textwidth}{(d)}
            \fig{joints1835.pdf}{0.33\textwidth}{(e)}}
\caption{Upper panel: Best fit final model with data. Lower panel: data-model/error. }
\label{fig:joint_eeudelc}
\end{figure*}
\end{comment}

\begin{deluxetable*}{lcccccccccc}
\tablenum{4}
\tablecaption{Model integrated unabsorbed continuum fluxes of different emission components based on the \opt{}/\fagn{} model. All the fluxes are in units of \ergS. Bolometric luminosities are calculated from the model integrated fluxes.}\label{tab:flux_comp_opt_fagn}
\tablewidth{0pt}
\tablehead{
Sources & Mass  & Distance & Model & $F_{disk}$  & $F_{soft}$  &$F_{hard}$ & $\frac{L_{Bol}}{L_{Edd}}$& $\rm \frac{L_{0.001-0.1~keV}}{L_{bol}(0.001-100~keV)}$
\\
 & &  & &$0.001-0.01$ keV &$0.5-2$ keV & $2-10$ keV&
 \\
 & ($10^8$\msun)& (Mpc) & & ($10^{-10}$) & ($10^{-11}$) &($10^{-11}$) & }
\startdata
%MR~2251   &  &  &   \\
PG0804 & 5.4$^a$ & 447.5& \opt{}&$1.85$ & $0.1$ & $1.0$ & 0.54& 0.63\\
&&&\fagn{}& $0.88$ & $0.2$ & $1.0$ & $0.20$ & $0.14$ \\
NGC~7469 & 0.1$^b$ & 68.7& \opt{}&$1.39$ & $13.9$  &$3.8$ & 0.72 & 0.32   \\%for Gamma 1.8
&&&\fagn{} & $1.17$ & $17.2$ & $3.9$ & $0.72$ &$0.07$ \\
SWIFT1921 & 0.39$^c$ &158 & \opt{} &$0.29$ & $1.4$ &$5.2$ & $0.56$ & $0.03$ \\
&&&\fagn{}& $1.01$ & $1.4$ & $5.2$ & $0.98$ & $0.06$ \\
SWIFT1835 & 10$^d$& 233.8 & \opt{} & $0.5$ & $0.2$ & $2.7$ & $0.01$& $0.2$\\
&&&\fagn{} & $0.7$ & $0.2$ & $2.7$ & $0.01$ & $0.2$ \\
\enddata
\tablecomments{Black hole masses are taken from: a -- \citealt{2015PASP..127...67B}; b --\citealt{2004ApJ...613..682P}; c --\citealt{Wang_2007}; d --\citealt{10.1111/j.1365-2966.2004.07822.x}}
\end{deluxetable*}

 \section{Results and Discussion}
 \label{sec:results}
We analyzed the simultaneous UV -- X-ray spectra of four type~1 AGN observed with \asat{}. We used the model \opt{} and \fagn{} to fit the broadband SEDs for all four sources. The soft X-ray excess emission below 2~keV is well described by warm Comptonization of the disk seed photons.  We obtain the  X-ray power-law photon index in the range of $\sim 1.5-2.1$. The inner accretion disk appears to be converted into warm Comptonizing plasma in all the sources. The disk ($0.001-0.01$ keV), soft X-ray ($0.5-2$ keV), and hard X-ray ($2-10$ keV) fluxes are listed in Table~\ref{tab:flux_comp_opt_fagn}.

%The black hole masses adopted for these four objects are  $10^7$\msun (NGC~7469; \citealt{2004ApJ...613..682P}), $5.4\times 10^8$\msun (PG0804; \citealt{2015PASP..127...67B}), $1.25 \times 10^7$\msun (SWIFT1921; \citealt{Wang_2007}) and $10^9$\msun (SWIFT1835; \citealt{10.1111/j.1365-2966.2004.07822.x}).

\subsection{PG~0804$+$761}
\label{subsec:pg08}

The accretion flow geometry was better described by the \fagn{} model, an outer standard disk, inner warm corona, and hot corona. In this case, the soft X-ray excess emission in the $0.2-2$ keV band is dominated by the thermal Comptonization of disk seed photons, while in \opt{}, a fraction of the soft excess is contributed by color-corrected disk blackbody emission. The difference in the underlying geometry and the assumptions between these two models may be the reason behind this difference. We obtained the Eddington ratio $\sim 0.2-0.5$, similar to that obtained by \citet{2018A&A...611A..59P} (Eddington ratio $\sim 0.4$).  
%Since the \fagn{} model does not use color correction, the \opt{} may be preferred despite the somewhat lower fit statistic.

%MR 2251-178: Earlier observation of MR 2251
\subsection{NGC~7469}
\label{subsec:ngc7469}
Based on our UV/X-ray SED, we estimated the bolometric luminosity $\sim 9 \times 10^{44}$~\Lum{}, corresponding to an Eddington ratio of $0.7$. During the observation performed in 1996 with \textit{IUE}/XTE, \citet{2004A&A...413..477P} found the bolometric luminosity, $L_{bol} \sim 2-3 \times 10^{44}$~\Lum{} which is lower than that obtained during our observation. Our joint UV/X-ray spectral modeling suggests that the accretion disk emits like a standard disk down to $r_{cor} \sim 39~r_g$, while the inner disk is transformed into a warm corona.  This is consistent with our UVIT/grating spectral analysis in paper~I, where we found the standard accretion disk to be truncated at $35-115~r_g$. 
%The discrepancy could be explained in the following way. The UVIT spectrum alone was insufficient to provide information about the contribution of soft X-ray excess and the X-ray power-law emission. Therefore, the \opt{} model may have represented the outer standard disk only, where the local effective peak temperature is much lower ($\sim 4.4$~eV), resulting in a large $r_{cor}$. In the broadband UV/X-ray modeling, to produce the observed soft X-ray excess, the inner edge of the disk is required to be hotter than that observed with UVIT-only data. Since the seed photons for the warm Comptonization are extracted from the edge at $r_{cor}$, the color-corrected disk extends till $\sim 14.5 ~r_g$, which could generate hot enough seed photons. The optical depth \rev{($\tau \sim  13.3$)} of the warm plasma is consistent with that obtained during 2015 \xmm{} observations \citep{middei2018multi}.  

\begin{deluxetable}{lcccc}
\tablenum{5}
\tablecaption{NGC 7469: Intrinsic luminosities (in units of $10^{43}$~\Lum) of different emission components in NGC~7469.  The luminosities in 2002 and 2015 are quoted from \xmm{} observations \citet{mehdipour2018multi}, while those in 2017 are derived in this work. \label{tab:7469_lum_comp}}
\tablewidth{0pt}
\tablehead{
\multirow{2}{*}{\begin{tabular}{c} Year\end{tabular}} & \colhead{$L_{disk}$ } & \colhead{ $L_{soft}$} &$L_{hard}$
\\
 & \colhead{($1000-7000$ \ang{})} & \colhead{($0.2-2$ keV)} &\colhead{($2-10$ keV)}}
\startdata
%MR~2251   &  &  &   \\
2002 & $6.3$ & $1$ & $3.5$\\
2015 &  $5.5$ & $0.8$ &$4.7$   \\
2017 &  $8.3$ & $21.5$ &$2.0$ \\
\enddata
\end{deluxetable}

\citet{mehdipour2018multi} modeled the broadband data acquired with \textit{SWIFT}-UVOT, \textit{HST}, and \textit{Chandra} using disk blackbody, warm Comptonization, hard X-ray power law, and a reflection model in two epochs, 2002 and 2015. 
%They found that the UV/optical luminosity ($1000-7000$ \ang{}) dropped from $\sim 6.3 \times 10^{43}$~\Lum~ in 2002 to $\sim 5.5 \times 10^{43}$~\Lum~ in 2015 (see Table \ref{tab:7469_lum_comp}). The soft X-ray ($0.2-2$~keV) luminosity showed a similar trend, dropping from $\sim 10^{43}$~\Lum~ (2002) to  $\sim 8.2 \times 10^{42}$~\Lum~ (2015). On the other hand, the $2-10\kev$ luminosity of the X-ray power-law component seemed rather uncorrelated with the soft X-ray component. It increased from $3.5 \times 10^{43}$~\Lum~ (2002) to $4.7 \times 10^{43}$~\Lum~ (2015). The luminosities during our observation in 2017 were $7.4 \times 10^{43}$~\Lum ~($1000-7000$ \angs{}), $2.5 \times 10^{43}$ \Lum ~($0.2-2$ keV), and $3.3 \times 10^{43}$ \Lum ($2-10$~keV). 
They found that both the UV/optical and soft X-ray luminosity followed a similar trend while the hard X-ray luminosity appear to be uncorrelated in 2002 and 2015 (see Table~\ref{tab:7469_lum_comp}). Based on these two epochs of observations, they concluded the soft X-ray excess may favor the warm Comptonization. Adding to that, with our \asat{} observation, we found the trend to be consistent with that found by \citet{mehdipour2018multi}. Apparently, in these three epochs, the UV and soft X-ray show a similar trend, favoring the warm Comptonization model as the origin of soft excess. 
%\changes{We tabulated the luminosities of three epochs in Table~\ref{tab:7469_lum_comp}.}

%A similar value was obtained in 2002 ($3\times10^{44}$\Lum) and 2015 ($3.8 \times 10^{44}$\Lum) by \citep{mehdipour2018multi}.    

\subsection{SWIFT~J1921.1$-$5842}
\label{subsec:swift1921}

\begin{figure}
\centering
  \includegraphics[scale=0.35]{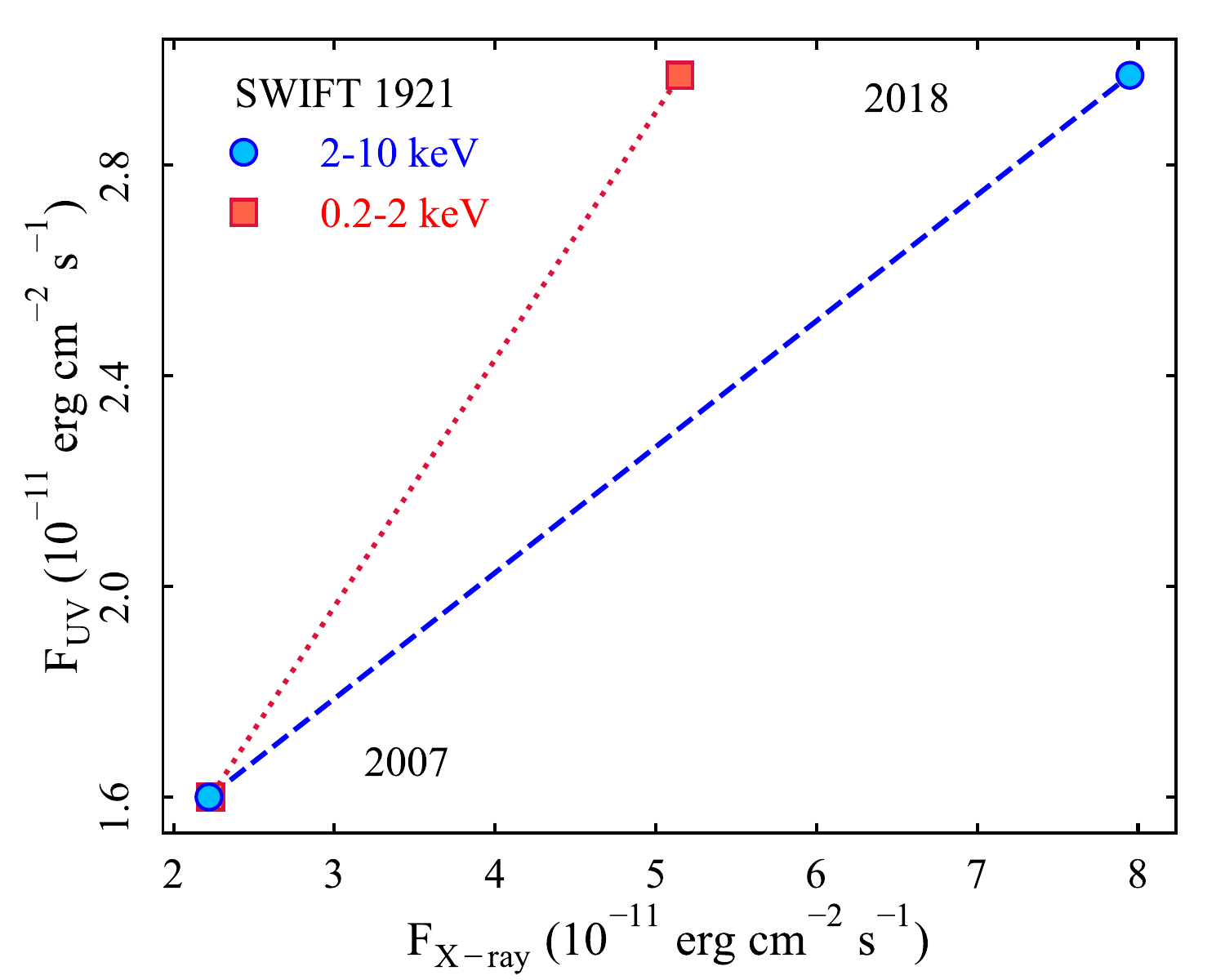}
\caption{Unabsorbed flux variation in 2007 \citep{ghosh2020broad} and 2018 (\asat{}). The $F_{UV}$ is the model integrated Galactic absorption corrected flux in the wavelength range $1870-2370$\ang{}.}
\label{fig:s1921_flx}
\end{figure}

The UVIT/grating analysis with \dbb{} model predicted high inner disk temperature ($kT_{in}>23$~eV), and the \opt{} model resulted in a poor quality fit ($\chi^2/dof = 492/321$) for a maximally rotating black hole (see paper~I). These results based on \dbb{} and \opt{}   may indicate that the UVIT/grating energy band in SWIFT1921 is significantly contributed by high-energy thermal Comptonized disk photons rather than standard or color-corrected accretion disk emission. The UV/X-ray spectral analysis resulted in a large $r_{cor} \sim 95 ~r_g$. Apparently, a large fraction of the BBB is contributed by the Comptonized photons. A similar scenario is observed in Fairall 9 by \citet{hagen2023estimating}.

  \begin{deluxetable}{ccccc}
\tablenum{6}
\tablecaption{SWIFT1921: The fluxes (in the unit of $10^{-11}$\ergS) in the 2007 are adopted from \citet{ghosh2020broad}, and those in 2018 are from this work. The first column lists the Galactic extinction corrected flux in the \xmm/OM-UVW2 filter (2007; adopted from \citet{ghosh2020broad}).  We calculated the $F_{UV}$ (2018) by integrating the extinction-corrected best-fit model flux in the UVW2 waveband. The last two columns represent the unabsorbed continuum fluxes of the soft and hard X-ray emission in SWIFT1921. \label{tab:s1921_flux_comp}}
\tablewidth{0pt}
\tablehead{
\multirow{2}{*}{\begin{tabular}{c}Observation \\year\end{tabular}} & \colhead{ $F_{UV} $} & $F_{soft}$ & $F_{hard}$
\\
 & \colhead{($1870-2370$ \ang{})} & \colhead{($0.2-2$ keV)} &\colhead{($2-10$ keV)}}
\startdata
%MR~2251   &  &  &   \\
2007 & $1.60$ & $2.22$ & $2.23$\\
2018 &  $2.97$ & $7.95$ &$5.15$   \\
\enddata
\end{deluxetable}

%In the broadband modeling, we obtained the unabsorbed $2-10$ keV flux to be $\sim 5.15 \times 10^{-11}$\ergS. This is a factor of two higher than those found in four sets of \xmm~ observations during  2007 ($\sim 2.2 \times 10^{-11}$\ergS; \citealt{ghosh2020broad}) although the spectral shape is similar during these two epochs. Also the soft X-ray flux in the $0.2-2$ keV band is increased from $2.2 \times 10^{-11}$\ergS ~(2007) to $4.4 \times 10^{-11}$\ergS. 
In Table \ref{tab:s1921_flux_comp}, $F_{UV}$ in 2007 represents the Galactic reddening corrected average UVW2 filter flux ($3.3 \times 10^{-14}$\ergA) for the three \xmm{} observations during October 2007. The total flux in the 50 nm wide band ($5.2-6.6$~eV) will be $1.6 \times 10^{-11}$\ergS. 
%From our best-fit model,  we found the model integrated Galactic reddening corrected flux $3 \times 10^{-11}$\ergS{} in the $5 - 6.6$~eV. 
It can be seen from Fig.~\ref{fig:s1921_flx}, the flux in all three components (UV, soft X-ray, and hard X-ray) during our observation has increased by a factor of $\sim 2-4$ (see also Table~\ref{tab:s1921_flux_comp}). The Eddington ratio in our observation is higher ($L_{bol}/L_{Edd} \sim 0.6-1$) than that inferred ($L_{bol}/L_{Edd} \sim 0.12-0.42$) from the broadband X-ray spectral modeling by \citet{ghosh2020broad}. 
Additionally, they found the soft X-ray and the Galactic extinction corrected UV flux measured at 2120 \angs{} (UVW2 filter) to be uncorrelated. 

The large inner disk radius obtained in our UV/X-ray spectral fitting may favor the warm Comptonization as the soft excess emission.  However, based on the broadband X-ray modeling with \xmm-\nustar, \citet{ghosh2020broad} concluded that both the warm Comptonization and the blurred reflection model describe the soft excess well. 
\citet{gondoin2003xmm} found a spectral slope of $\Gamma \sim 1.78$ and the $2-10$~keV absorbed flux $\sim 2.65 \times 10^{-11}$\ergS~ using the \xmm~ observation performed in 2001. We found a steeper $\Gamma \sim 2$ and higher $2-10$~keV absorbed flux $\sim 3.3 \times 10^{-11}$\ergS~, which may indicate the thermal (hot) Comptonization being responsible for the UV - hard X-ray spectral variability in this source.

%\boldmath
  \begin{deluxetable}{ccccc}
\tablenum{7}
\tablecaption{SWIFT1835: All the values are in the unit of \ergS. The first two rows show the highest (O5) and the lowest fluxes (O1) observed by \citet{ursini2018radio} in five epochs of observation during 2016 using the data acquired with  \xmm/\nustar{}. The model integrated fluxes in the last row are obtained from \asat{} observation.
\label{tab:s1835_thermal_comp}}
\tablewidth{0pt}
\tablehead{
\multirow{2}{*}{\begin{tabular}{c}\textbf{Observation} \\\textbf{year}\end{tabular}} 
 & \colhead{$F_{5.2-6.6 ~eV}$} & \colhead{$F_{0.3-2~keV}$} &\colhead{$F_{2-10~keV}$}\\
 & \colhead{ ($10^{-12}$) } & ($10^{-12}$) &($10^{-11}$)}
\startdata
%2016   &  &  &   \\
2016(O1) & $6.1$ & $8$ & $3.4$\\
2016(O5) &  $7.5$ & $18$ &$4.2$    \\
2018 & $4.9$&$5.7$ &$2.8$ \\
\enddata
\end{deluxetable}

  \begin{deluxetable}{ccccccccc}
\tablenum{8}
\tablecaption{SWIFT1835: The model expression in XSPEC: \texttt{TBABS $\times$ XABS $\times$ REDDEN $\times$ (\nthc$^h$ + \nthc$^w$ + GAUSSIAN+GAUSSIAN+GAUSSIAN)}. \label{tab:s1835_nth2}}
\tablewidth{0pt}
\tablehead{
 $\Gamma_w$ & $kT_w$ & $kT_{disk}$ &$Norm_w$& $\Gamma_h$ & $Norm_h$
\\
  & (keV) & (eV) & ($10^{-4}$) & & ($10^{-3}$) }
\startdata
 $2.2_{-0.2}^{+0.2}$& $0.16_{-0.09}^{+0.07}$ & $<1.45$ & $7.2_{-4.1}^{+6.7}$ & $1.56_{-0.09}^{+0.07}$ & $5.3_{-0.6}^{+0.4}$ \\
\enddata
\end{deluxetable}

\subsection{SWIFT~J1835.0$+$3240}
\label{subsec:swift1835}
The $r_{cor}$ inferred from our broadband spectral modeling is consistent with that derived using the UVIT/grating spectral analysis in paper~I.  
We obtained the $2-10$ keV flux, $\sim 2.8 \times 10^{-11}$\ergS~ and the X-ray power-law slope, $\Gamma = 1.56_{-0.09}^{+0.07}$.  
%\citet{ballantyne2014nustar} found, the $2-10$~ keV flux  $\sim 5 \times 10^{-11}$\ergS~ in 2012 and   $2.9 \times 10^{-11}$\ergS~ in 2013 observations. They classified these as the high and low flux states, respectively. The photon indices are $1.68_{-0.02}^{+0.03}$ and $1.78_{-0.03}^{+0.02}$ in the low and high flux states, respectively. They conclude that the bright flux state exhibiting steeper spectra results from thermal Comptonization.
Utilizing the data acquired with \xmm{}/OM filters (U, UVW1, UVW2, and UVM2), EPIC-pn and \nustar{} in 2016, \citet{ursini2018radio} modeled their UV -- X-ray spectra with \ion{Fe}{2} and Balmer continuum (small blue bump in UV), two thermal Comptonization model (warm and hot), one warm absorber, and two emission lines (in X-ray). They found the $2-10$~keV flux varied in the range $3.4-4.2 \times 10^{-11}$\ergS (Table \ref{tab:s1835_thermal_comp}), while the $\Gamma_h$ remained fairly constant at $\sim 1.8$.  
 %Additionally, we used similar model components as by  \citet{ursini2018radio} and modelled the data with two thermal Comptonization models and one warm absorber component with the parameter values fixed as obtained by them. 
By modeling our UV/X-ray spectra with two thermal Comptonization models and one warm absorber component, we found a lower ($kT_{disk}<1.45$~eV) disk temperature than that obtained by \citet{ursini2018radio} ($\sim 3.4$~eV). %Therefore, the hardening of the $\Gamma_h$ (see Table~\ref{tab:s1835_nth2} for other parameters) during our observation indicates that the thermal Comptonization process could be driving the UV/X-ray spectral variability. 
In addition, we found the accretion rate ($L_{Bol}/L_{Edd} \sim 0.01$) to be 50 -- 60\% lower than that observed during 2016 by \citet{ursini2018radio}. The low accretion rate coupled with a harder photon index compared to the previous observation, may indicate a transition from a bright/soft state to a dim/hard state. A similar spectral hardening has been observed by \citet{ballantyne2014nustar}. Based on two epochs of \nustar{} observations in 2007 and 2008, they observed a higher coronal temperature and harder $\Gamma$ in the low X-ray flux state ($2-10$ keV) compared to those in the high flux state. They concluded that the observed coronal heating is the consequence of Comptonization of disk seed photons, typically observed in many Seyfert galaxies \citep{dewangan200210,zdziarski2003correlations,tripathi2021revealing}.
%They concluded that SWIFT1835 exhibits a Seyfert-like Comptonizing corona   
 
 We found the $0.3-2$~keV flux and the electron temperature of the warm plasma ($kT_w \sim 0.14$~keV) lower (see Table~\ref{tab:s1835_thermal_comp}) than those obtained by \citealt{ursini2018radio} ($kT_w \sim 0.5$~keV). Therefore, the plasma temperature in the warm corona has reduced with the reduction in overall flux in the UV to hard X-ray band (see Table~\ref{tab:s1835_thermal_comp}), indicating this to be associated with the change in the accretion rate.  
 
% This hardening of the $\Gamma$ can be explained by considering the outflowing corona. As explained by \citet{ballantyne2014nustar}, from the model of \citet{malzac2001x}, outflow velocity of $>0.5c$ would result in $\Gamma \sim1.7$ and a reflection fraction $R<0.2$. 
 %If the corona is outflowing, lesser amount of the disk emission will intercept the corona and vice-versa. 
% As noted by \citet{ballantyne2014nustar, ursini2018radio}, outflowing corona also explains the weak X-ray reflection observed in the BLRGs.  This may indicate a connection of corona with the radio jet, the base of which is believed to be the corona \citep{markoff2005going}.  
 
 %Additionally, we found that the hard X-ray and the soft X-ray flux both reduced during our observation. This is consistent with the trend observed in 2016 by \citet{ursini2018radio}. Eddington ratio during our observation is consistent with \citet{ursini2018radio}. 

\section{Conclusion}
\label{sec:conclusion}

We present the UV -- X-ray broadband spectral analysis of four type~1 AGN: PG0804, NGC~7469, SWIFT1921, and SWIFT1835 utilizing \astrosat{} observations. We found the soft excess to be consistent with the warm Comptonization in these sources. The main results of our SED modeling are described below:

\begin{enumerate}

    \item PG0804 shows little to no flux variation in the emission components compared to the previous observations in 2010 with \xmm~ \citep{2018A&A...611A..59P}. 
    %The Eddington ratio ($0.4-0.5$) is also similar in these observations. 
   Our UV/X-ray data are better described by a standard outer disk and inner warm and hot corona. 
   We obtained the spin parameter of $0.76_{-0.20}^{+0.08}$ (1$\sigma$ error) with the \fagn{}.
    %We found the inner disk radius extends down to $\sim 1.5~r_g$ (near ISCO), \rev{and a relatively steeper index ($\Gamma_{warm }\sim 3.2$) for the warm plasma. These results may indicate that the soft X-ray excess emission has some contribution from the X-ray reflection.}
    
    \item We found that NGC~7469 favors the warm Comptonization scenario for the origin of soft excess. This source appears to exhibit low to moderate black hole spin ($a^\star<0.67$). 
    %We obtained an upper limit of 0.5 in the black hole spin.

    \item For SWIFT1921, we found that the fluxes in all three components, UV, soft X-ray excess, and X-ray power-law, are twice higher than \citet{ghosh2020broad} during our observation. The standard disk appears truncated at a large radius of $95 ~r_g$.  The hard X-ray spectral slope is consistent with \citet{ghosh2020broad}. 

    \item In the case of SWIFT1835, both the UV/optical and X-ray fluxes decreased during our observation compared to the \xmm~ observations by \citet{ursini2018radio}. Their X-ray power-law photon index ($\Gamma \sim 1.78$) and Eddington ratio ($0.02-0.03$) differ from those during our observation ($\Gamma \sim 1.56$, $L_{Bol}/L_{Edd}\sim 0.01$). The hardening of $\Gamma$ with the reduction in the disk temperature and accretion rate may indicate a state transition from a high/soft to a google map low/hard state in this source. 
\end{enumerate}

\begin{acknowledgments}
This publication uses data from Indian
Space Science Data Centre (ISSDC) of the \textit{AstroSat} mission of the Indian Space Research Organisation (ISRO) and \xmm.  We acknowledge the  SXT POC at TIFR (Mumbai) and UVIT POC at IIA (Bangalore)  for providing the necessary software tools for data processing. The UVIT data were processed by CCDLAB pipeline \citep{Postma_2017}. This research has used the Python and Julia packages. This research has used the SIMBAD/NED database. S.K. acknowledges the University Grant Commission (UGC),
Government of India, for financial support. K. P. Singh thanks the Indian National Science Academy for support under the INSA Senior Scientist Programme. L.M. acknowledges support from the CITA National Fellowship (reference \# DIS-2022-568580) Program.
\end{acknowledgments}

\vspace{5mm}
\facilities{\asat}

\software{XSPEC \citep{arnaud1996xspec}, SAOImageDS9 \citep{joye2003new}, Julia \citep{bezanson2017julia}, Astropy \citep{2013A&A...558A..33A,2018AJ....156..123A}}

\bibliography{uv_xray_joint_shrabani}{}
\bibliographystyle{aasjournal}

\end{document}